\newcommand{\CLASSINPUTtoptextmargin}{2.4cm}
\newcommand{\CLASSINPUTbottomtextmargin}{4.6cm}
\documentclass[conference]{IEEEtran}
\IEEEoverridecommandlockouts
\usepackage{authblk}
\usepackage{cite}
\usepackage{amsmath,amssymb,amsfonts}
\usepackage{algorithmicx}
\usepackage{algorithm,algpseudocode}
\usepackage{graphicx}
\usepackage{textcomp}
\usepackage{xcolor}
\usepackage{tabularx}
\usepackage{url}
\usepackage[bookmarks=false]{hyperref}
\usepackage{rotating}
\usepackage{subcaption}
\usepackage{multirow}
\usepackage{pifont}
\usepackage{tcolorbox}
\definecolor{tumColorLightBlue}{HTML}{f0f5fa} 
\usepackage{cleveref}

\usepackage{tablefootnote}
\usepackage{threeparttable}
\usepackage{multirow}
\usepackage{booktabs}
\usepackage{harveyballs}
\renewcommand{\baselinestretch}{0.96}
\usepackage{eso-pic}
\usepackage[outdir=./]{epstopdf}
\usepackage{fancyhdr}

\def\BibTeX{{\rm B\kern-.05em{\sc i\kern-.025em b}\kern-.08em
    T\kern-.1667em\lower.7ex\hbox{E}\kern-.125emX}}

\AddToShipoutPictureBG*{%
  \AtPageUpperLeft{%
    \put(0.07\paperwidth,-1.2cm){%
      \fbox{%
        \begin{minipage}{\textwidth}
          \centering
          \small
          \textbf{This paper has been accepted for publication in the proceedings of the IEEE Consumer Communications \& Networking Conference (CCNC), January 2025. The copyright for this paper will be transferred to IEEE. ©2023 IEEE. Personal use of this material is permitted. Permission from IEEE must be obtained for all other uses, in any current or future media, including reprinting/republishing this material for advertising or promotional purposes, creating new collective works, for resale or redistribution to servers or lists, or reuse of any copyrighted component of this work in other works.}
        \end{minipage}
      }%
    }%
  }%
}

\begin{document}

\title{CyclicSim: Comprehensive Evaluation of Cyclic Shapers in Time-Sensitive Networking\vspace{-0.3cm}}

\author[1]{Rubi Debnath}
\author[2]{Luxi Zhao}
\author[3]{Mohammadreza Barzegaran}
\author[4]{Sebastian Steinhorst\vspace{-0.2cm}} 

\affil[1,4]{TUM School of Computation, Information and Technology, Technical University of Munich, Germany}
\affil[2]{Department of Electronic and Information Engineering, Beihang University, Beijing, China}
\affil[3]{Department of Electrical Engineering and Computer Science, University of California, Irvine, USA}
{
    \makeatletter
    \renewcommand\AB@affilsepx{, \protect\Affilfont}
    \makeatother
    \affil[1,4]{firstname.lastname@tum.de}
    \affil[2]{zhaoluxi@buaa.edu.cn}
    \affil[3]{barzegm1@uci.edu}
}

\makeatletter
\patchcmd{\@maketitle}
  {\addvspace{0.5\baselineskip}\egroup}
  {\addvspace{-1\baselineskip}\egroup}
  {}
  {}
\makeatother

\maketitle

\begin{abstract}
Cyclic Queuing and Forwarding (CQF) is a key Time-Sensitive Networking (TSN) shaping mechanism that ensures bounded latency using a simple gate control list (GCL). Recently, variants of CQF, including Cycle Specific Queuing and Forwarding (CSQF) and Multi Cyclic Queuing and Forwarding (MCQF), have emerged. While popular TSN mechanisms such as the Time-Aware Shaper (TAS), Asynchronous Traffic Shaper (ATS), Credit-Based Shaper (CBS), and Strict Priority (SP) have been extensively studied, cyclic shapers have not been thoroughly evaluated. This paper presents a comprehensive analysis of CQF, CSQF, and MCQF, providing insights into their performance. We quantify delays through simulations and quantitative analysis on both synthetic and realistic networks. For the first time, we introduce an open-source OMNeT++ and INET4.4 based framework capable of modeling all three cyclic shaper variants. Our tool facilitates the validation of new algorithms and serves as a benchmark for cyclic shapers. Our evaluations reveal that MCQF supports diverse timing requirements, whereas CSQF, with its additional queue, often results in larger delays and jitter for some TT flows compared to CQF. Additionally, CSQF does not demonstrate significant advantages in TSN networks where propagation delays are less critical than in wide-area networks (WANs). 
\end{abstract}

\section{Introduction}
\label{sec:introduction}
Time-Sensitive Networking (TSN)\cite{8021Q, cqf_survey_ahmed} is revolutionizing not only the industrial domain but also the aviation, vehicular, and spacecraft sectors by providing deterministic communication for layer-2 networks. TSN, with its further extension into the wireless domain such as 5G-TSN\cite{rubi_vtc} and WiFi-TSN targets deterministic communication for wireless movable robot arms and devices in cross-domain networks. As the need for timing guarantees grows, so does the demand for fast and scalable algorithms, precise verification methods, and simulation-based verification tools. The IEEE 802.1 TSN Task Group (TG) is working toward developing a TSN suite that includes sub-standards for IEEE 802.1 Ethernet. As a result, TSN offers multiple sub-standards featuring various shaping and scheduling mechanisms that provide different quality of service (QoS) levels, such as the Time-Aware Shaper (TAS)\cite{8021Qbv}, Asynchronous Traffic Shaper (ATS)\cite{8021Qcr}, Credit-Based Shaper (CBS)\cite{8021QAV}, and Cyclic Queuing and Forwarding (CQF)\cite{8021Qch}.

As the most popular mechanism in TSN technology, the TAS scheduling mechanism has been extensively studied over the years. TAS provides timing guarantees and zero jitter by precisely \texttt{opening} and \texttt{closing} gates for Time-Triggered (TT) traffic types. This gate mechanism is referred to as the Gate Control List (GCL). However, generating the GCL for TAS is an NP-hard problem, and its complexity scales with the network size and the number of flows. Consequently, related work~\cite{luxi_tnsm, arxiv_sch_survey, rubi_rtcsa} has proposed using various shaping mechanisms, such as CBS, ATS, and CQF, in mixed-criticality networks to alleviate the load on TAS.

\begin{figure}[t!]
    \centering    
    \includegraphics[scale=0.21, trim={0.7cm 0.5cm 0.7cm 0.5cm}, clip]{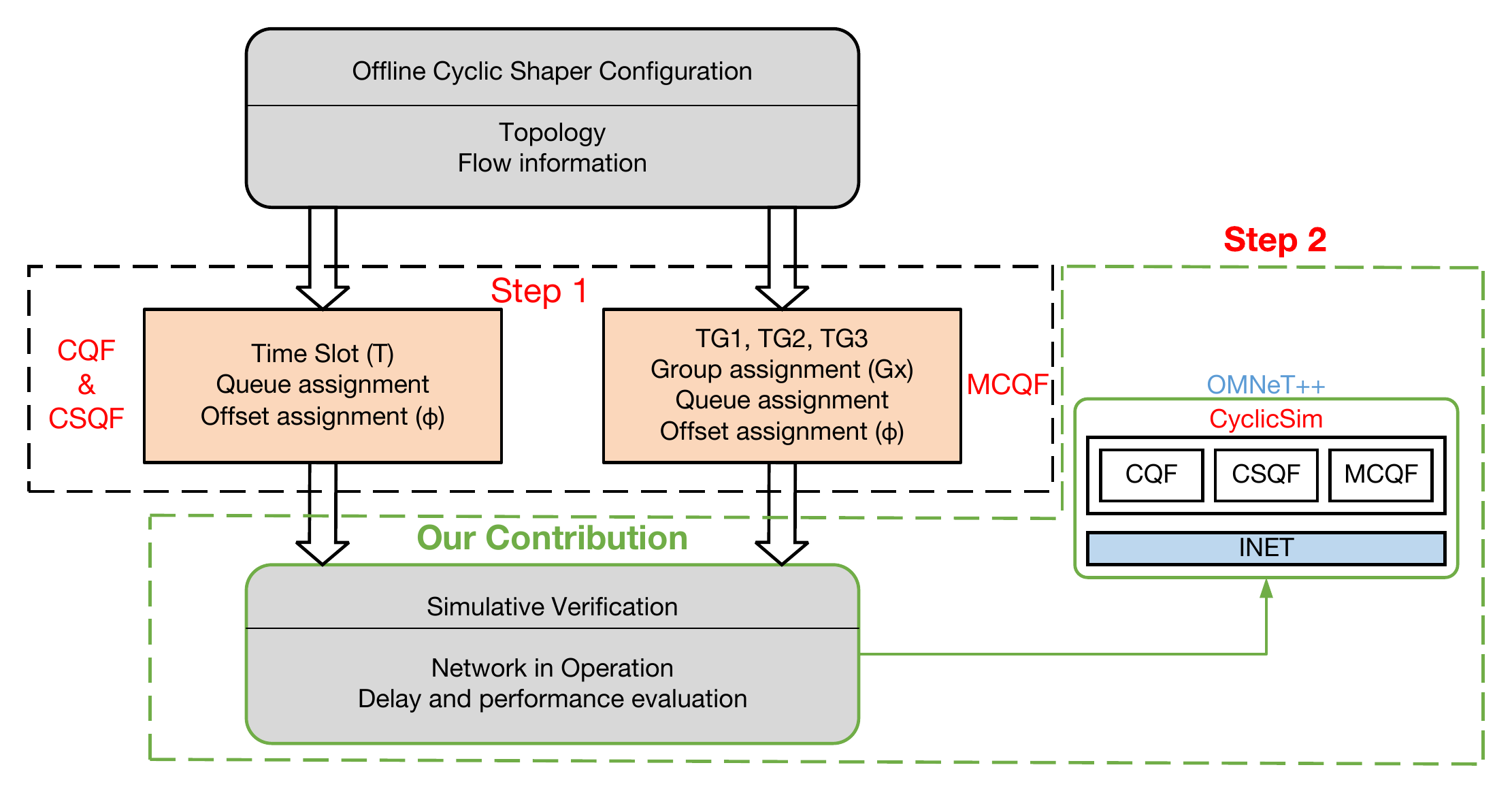}
    \caption{Overview of the cyclic shapers implementation and verification highlighting our contribution.}
\label{fig:ccnc_overview}
\vspace{-0.6cm}
\end{figure}

\begin{figure*}[t]
    \centering
    \begin{minipage}[b]{0.32\textwidth}
        \centering
        \includegraphics[scale=0.1, trim={0cm 0cm 0cm 0cm}, clip]{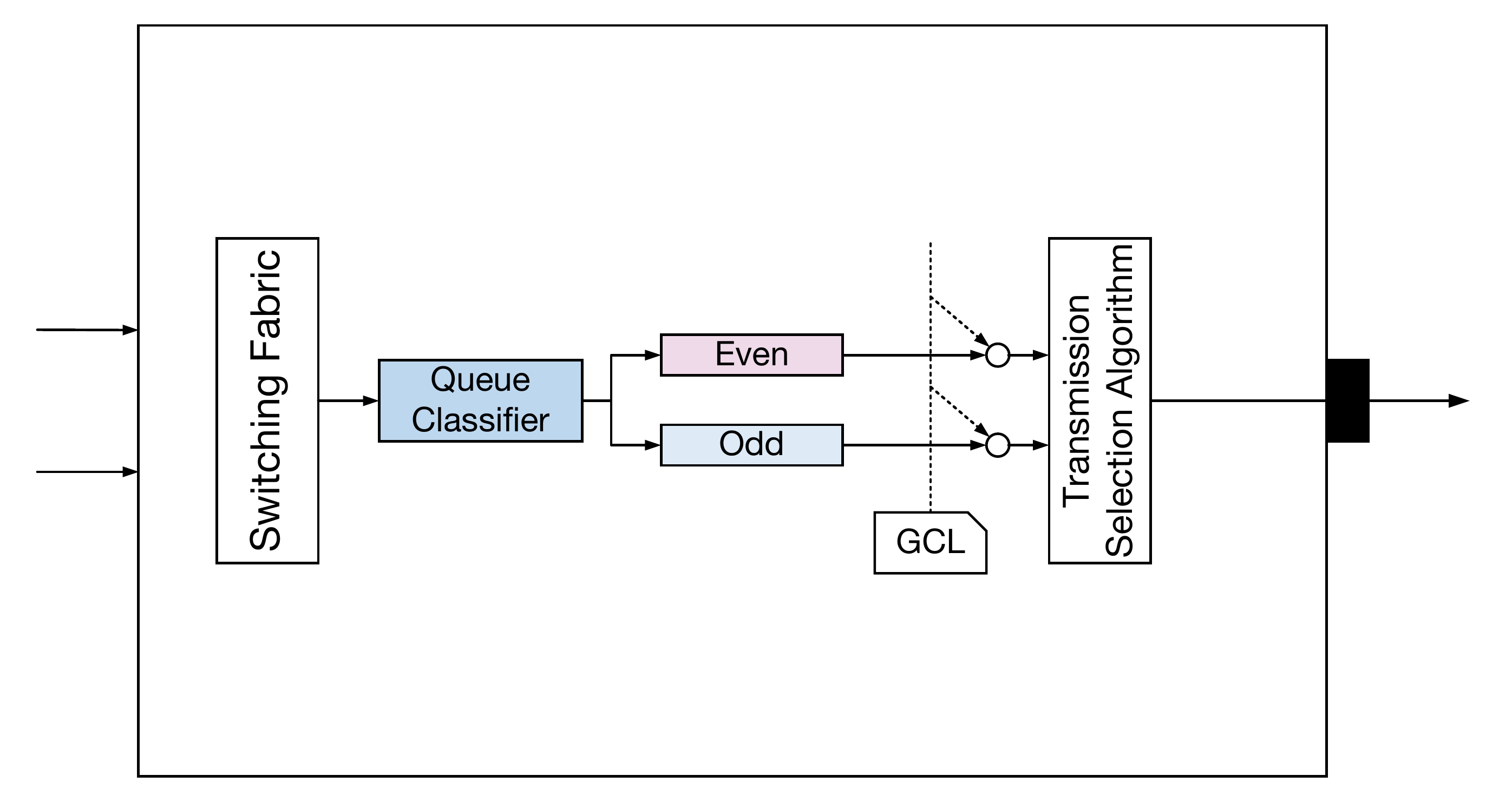}
        \subcaption[]{CQF architecture.}
    \label{fig:cqf_overview}
    \end{minipage}
    \hfill
    \begin{minipage}[b]{0.32\textwidth}
        \centering
        \includegraphics[scale=0.1, trim={0cm 0cm 0cm 0cm}, clip]{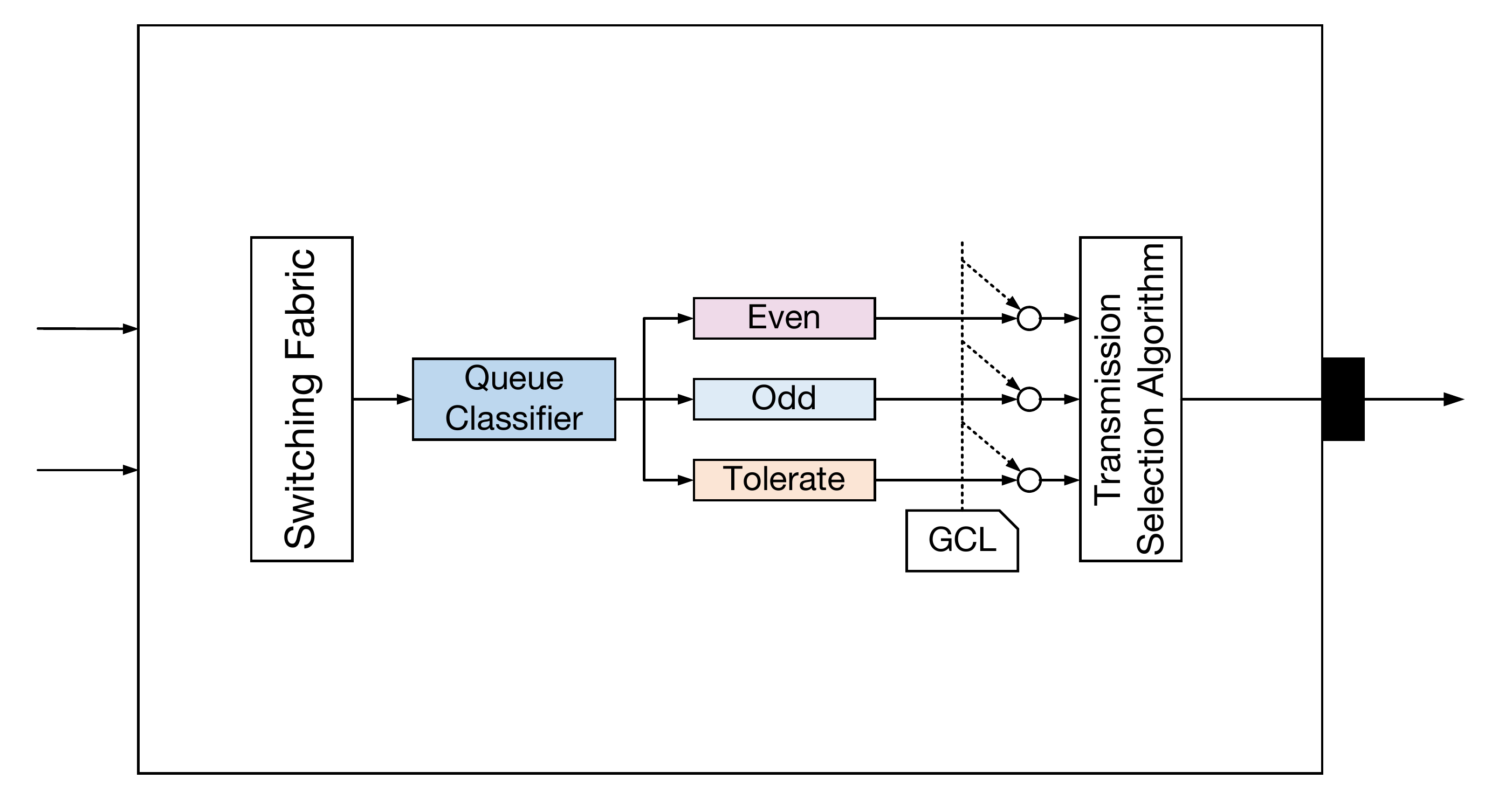}
        \subcaption[]{CSQF (3-queue CQF) architecture.}
    \label{fig:csqf_overview}
    \end{minipage}
    \hfill
    \begin{minipage}[b]{0.32\textwidth}
        \centering
        \includegraphics[scale=0.1, trim={0cm 0cm 0cm 0cm}, clip]{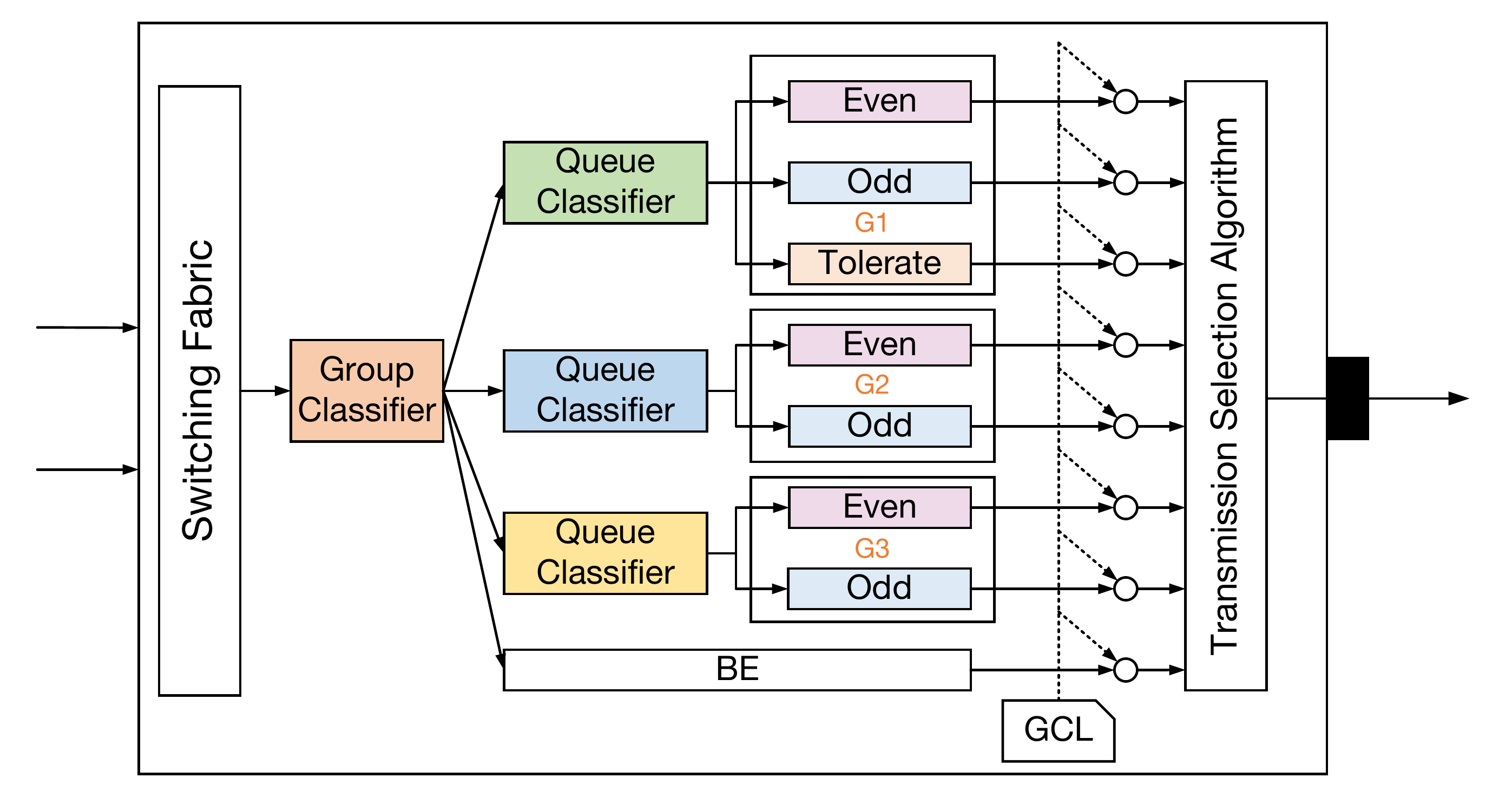}
        \subcaption[]{MCQF architecture.}
    \label{fig:mcqf_overview}
    \end{minipage}
    \caption{Architecture of different variants of cyclic shapers in TSN.}
    \label{fig:overall}
    \vspace{-0.5cm}
\end{figure*}

Like TAS, CQF is also based on the GCL. CQF is gaining popularity due to its simple GCL operation and straightforward Worst-Case Delay (WCD) calculation. Although CQF does not provide the fine-grained scheduling that TAS offers, it features easier scheduling configuration with bounded delay and jitter. Over the years, different variants of CQF have been proposed, such as Cycle Specific Queuing and Forwarding (CSQF)~\cite{ietfSegmentRouting} and Multi Cyclic Queuing and Forwarding (MCQF)\cite{norman,mcqf_paul}. Therefore, in a mixed-criticality network with diverse QoS requirements across various applications, selecting the appropriate scheduling and shaping algorithm has become increasingly challenging. The CQF standard does not specify the configuration and scheduling algorithm, leaving it to the research community to design and implement novel, efficient algorithms. These algorithms are often developed using mathematical solvers like satisfiability modulo theories (SMT) or SAT, or heuristics such as Simulated Annealing and Genetic Algorithms. However, the solutions generated by these methods are typically verified using Python or custom simulators, which lacks comprehensive TSN capabilities, potentially leading to errors and inaccuracies. Ideally, the verification of these algorithmic solutions should be conducted on simulators that accurately replicate real-world scenarios, such as OMNeT++, or directly on TSN hardware. 

Since MCQF is not yet standardized and cyclic shapers are still evolving, existing TSN switches lack hardware support for CQF, 3-queue CQF\footnote{In this paper, CSQF for TSN is referred to as \textbf{3-queue CQF} to distinguish it from the CSQF used in Layer-3 DetNet. Details are provided in Section~\ref{sub:csqf}.}, and MCQF. Thus, tools like OMNeT++ are crucial for identifying the benefits and drawbacks of new mechanisms. Simulation tools provide valuable insights into the performance and behavior of TSN mechanisms under various conditions, eliminating the need for expensive and complex physical hardware testbeds. As TSN research progresses, supporting all TSN mechanisms and their combinations becomes increasingly important~\cite{arxiv_sch_survey}. Although OMNeT++\footnote{\url{https://omnetpp.org/}} is widely used for TSN simulation~\cite{arxiv_sch_survey}, there is currently no open-source tool that can simulate all cyclic shaper variants. This need forms the core motivation behind our paper. 

\vspace{0.1cm}
\noindent\textbf{Motivation:}
The major roadmap for TSN cyclic shapers involves three critical steps: (i) designing the scheduling or configuration algorithm (shown as \textcolor{red}{Step 1} in Fig.~\ref{fig:ccnc_overview}), (ii) verifying the scheduling algorithm (shown as \textcolor{red}{Step 2} in Fig.~\ref{fig:ccnc_overview}), and (iii) conducting hardware evaluation. While Konstantinous et al. proposed scheduling algorithms for different variants of cyclic shapers in~\cite{mcqf_paul}, they relied on a Python-based simulator for verification. Similarly, other related works~\cite{itp_infocom, cqf_fp, mccqf_icc_yan, jrs_cqf_yang} also used custom Python or other software-based tools to verify their algorithms. This limitation has hindered comprehensive validation, universal benchmarking, and broader adoption of cyclic shapers. Although INET 4.4\footnote{\url{https://inet.omnetpp.org/2022-07-27-INET-4.4.1-released.html}} has implemented many TSN features, providing a common verification and simulation tool for everyone, it still lacks native support for cyclic shapers. To address this, we have developed an open-source OMNeT++ framework that supports all cyclic shaper variants, bridging the existing gap in simulation tools. Our framework enables the research community to validate cyclic shaper algorithms, eliminating the need for custom simulators and facilitating rigorous testing. This paper introduces our OMNeT++ and INET4.4 based simulation framework. Additionally, we share implementation-specific insights and key findings to further support the development and understanding of these mechanisms. In summary, our key contributions are:

\begin{figure*}[!t]
    \centering
        \includegraphics[scale=0.22, clip, trim={0cm 1cm 1cm 1cm}]{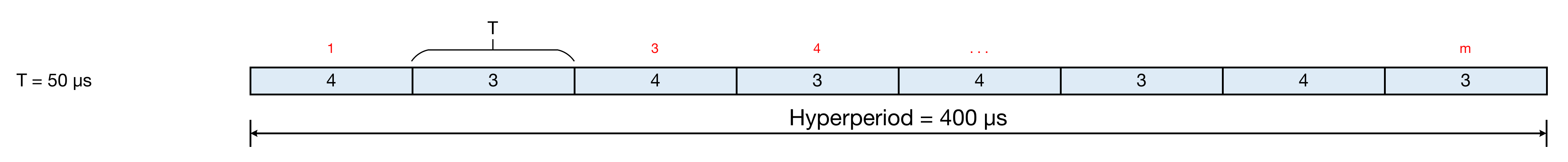}
        \caption{Hyperperiod Overview of the CQF and 3-queue CQF (CSQF) network operating with one time slot $T$.}
\label{fig:cqf_hyperperiod_example}
\vspace{-0.3cm}
\end{figure*}

\begin{figure*}[!t]
    \centering
        \includegraphics[scale=0.22, clip, trim={0cm 1cm 1cm 0cm}]{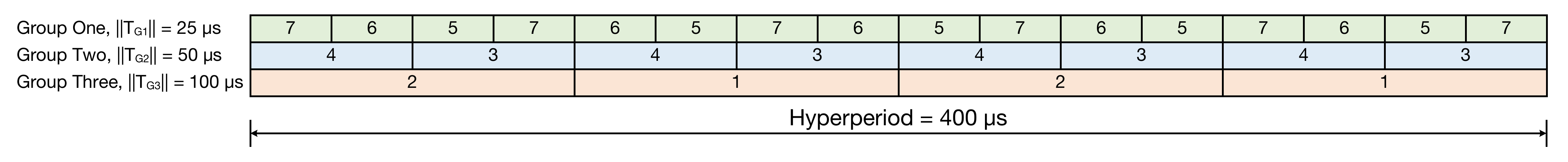}
        \caption{Hyperperiod overview of the MCQF for three \texttt{Groups} operating with three different time slots $T_{G1}$, $T_{G2}$, and $T_{G3}$.}
\label{fig:mcqf_hyperperiod_example}
\vspace{-0.42cm}
\end{figure*}

\begin{enumerate}
    \item \textbf{Cyclic Shaper Simulation:} We introduce an open-source simulator that supports different types of cyclic shapers in TSN, as detailed in Section~\ref{sec:cyclic_shapers}). 
    \item \textbf{Benchmarking Cyclic Shapers:} We provide configurations for various cyclic shapers to assist the scientific community in validating and benchmarking new algorithms. This enables effective verification and comparison, as discussed in Section~\ref{sec:system_model}. Additionally, we explore key findings and the rationale behind our implementation choices in Section~\ref{sec:results}.
    \item \textbf{Extensive Evaluation:} We conduct comprehensive experiments to evaluate the accuracy and performance of CQF, CSQF (3-queue CQF), and MCQF in both synthetic and realistic scenarios, as outlined in Section~\ref{sec:results}.
\end{enumerate}

\section{Cyclic Shapers}
\label{sec:cyclic_shapers}
In this section, we explore the different variants of cyclic shapers in TSN. We begin with an overview of CQF, followed by CSQF (3-queue CQF), and conclude with MCQF. Although these cyclic shapers share some common constraints and principles, each variant has its unique set of constraints. Therefore, we first describe the mechanisms and concepts of each cyclic shaper, before delving into their commonalities and specific constraints.
\vspace{-0.1cm}
\subsection{Cyclic Queuing and Forwarding}
\label{sub:cqf} 
Each TSN switch with CQF capability as shown in Fig.~\ref{fig:cqf_overview} has two queues for the CQF mechanism: \emph{even} and \emph{odd} queue, where in one time slot ($T$), one queue keeps receiving the incoming TT frames, and the another queue sends the frames received in the previous time slot. Time slot denoted as $T$ is the granularity of the scheduling in the cyclic shaper. The hypercycle or the scheduling cycle denoted as $H$ is divided into numbered time slots with a total of $m$ equal time slots (refer Fig.~\ref{fig:cqf_hyperperiod_example}). Each time slot is denoted as $T_j$, where $j$ is an integer number and $j\in \{1, \cdots, m\}$. Each time slot $T_j$ is of length $T$ and the length is given in $\mu$s. A sending node in the CQF network may transmit a frame during the time slot number $j$ to the next hop. The next intermediary switch receives the frame in the same time slot $T_j$, en-queues it and then forwards it to the next switch in time slot number $j+1$. The end-to-end delay of a TT flow from the sender to the receiver is determined exclusively by the time slot duration $T$ and number of switches ($SW_{num}$) in the routing of the TT flow. $SW_{num}$ is determined by the routing taken by the TT flow from the sender to the receiver. As the end-to-end delay in CQF is dependent on the time slot duration $T$, deciding on a proper $T$ value is crucial in a CQF network. A large time slot lead to high end-to-end delay, and small time slot requires high bandwidth (BW). Despite these challenges, CQF is well suited for real-time applications with loose requirements for delay and jitter boundaries~\cite{tamcqf_2024}. 

\noindent\textbf{WCD CQF:} The worst case end-to-end delay of the TT flows in the CQF network is quantified as follows:
\vspace{-0.1cm}
\begin{equation}
    \mathrm{Max \; Delay} = f_i.\phi + (\mathrm{SW_{num}}+1) \cdot T + \xi,
\end{equation} 
where $f_i\cdot \phi$ is the offset of the flow $f_i$ in $\mu$s, $\mathrm{SW_{num}}$ is the total number of switches in the route of the TT flow, $T$ is the time slot in $\mu$s, and $\xi$ denotes the network specific delays: processing delay, propagation delay, and time synchronization error ($\mathrm{sync_{error}}$).
\vspace{-0.1cm}
\begin{equation}
\label{eq:xi}
\xi = \mathrm{delay_{processing}} + \mathrm{delay_{propagation}} + \mathrm{sync_{error}}.
\end{equation}

\noindent\textbf{BCD CQF:} The best case end-to-end delay (BCD) of the TT flows in the CQF network is quantified as follows:
\vspace{-0.1cm}
\begin{equation}
    \mathrm{Min \; Delay} = f_i.\phi + (\mathrm{SW_{num}}-1) \cdot T + \xi.
\end{equation}

\noindent\textbf{Worst-Case Queuing Delay CQF:} The worst-case queuing delay for CQF is \textbf{zero} because the TT flows received by the switch in one time slot are transmitted in the next time slot.
 
\subsection{Cycle Specific Queuing and Forwarding (3-queue CQF)}
\label{sub:csqf} 
CSQF is an extension of CQF working with three-queues, where the third queue is an additional queue called the \emph{tolerating queue}. CSQF is proposed for the large scale deterministic network called DetNet~\cite{ietfSegmentRouting}. DetNet is used to provide deterministic communication in layer-3 over Wide Area Network (WAN). Unlike TSN, WAN has long-distance cables causing long propagation delays which causes the TT frames to sometimes arrive in the next time slot to the next node. Such frames which arrive late from its designated time slot are stored in the \emph{tolerating queue}. In CSQF, the sending time of all nodes along the path between the source and destination is known beforehand. Even though, CSQF is not standardized for TSN, there is a growing interest in CSQF within TSN due to the benefits provided by the \emph{tolerating queue}. This third queue enhances the schedulability of TT flows in TSN, making it crucial to study the performance of CSQF in TSN networks. In this paper, we implement CSQF for layer-2 TSN network. However, the implementation of CSQF in TSN is different from the layer-3 CSQF. Therefore, to avoid confusion, we hereafter refer the CSQF for TSN as \textbf{3-Queue CQF} as shown in Fig.~\ref{fig:csqf_overview}. Unlike CSQF, in  3-Queue CQF, we do not use the segment identifier (SID). In 3-Queue CQF, the TT frames are sent to the designated queue based on the Queue ID (\emph{qid}). TT flows stored in the \emph{tolerating queue} waits for one time slot before transmitting to the next node. Similar to CQF, in 3-queue CQF, the network operates with one time slot value ($T$) and $H$ is divided into $m$ equal time slots (as shown in Fig.~\ref{fig:cqf_hyperperiod_example}).

\noindent\textbf{WCD 3-queue CQF:} The worst case end-to-end delay of the TT flows in the 3-Queue CQF is defined as follows:
\vspace{-0.2cm}
\begin{equation}
    \mathrm{WCD} = f_i.\phi + (\mathrm{SW_{num}}+1)\cdot T + (\mathrm{SW_{num}} \cdot d_{queue}) + \xi,
    \vspace{-0.1cm}
\end{equation}
where $f_i\cdot \phi$ is the offset of the flow, $\mathrm{SW_{num}}$ is the total number of switches, $T$ is the time slot, $\xi$ denotes the network delays, and $d_{queue}$ is the queuing delay as explained below. 

\noindent\textbf{Worst-Case Queuing Delay 3-queue CQF:} A TT frame is delayed for a certain time slot in a 3-queue CQF network after being received by the switch. This delay is referred to as the queuing delay and is denoted by~$d_{queue}$. If the TT flow is stored in the \emph{tolerating} queue, it waits for one time slot. The worst-case queuing delay for 3-queue CQF is given below: 
\vspace{-0.1cm}
\begin{equation}
\label{eq:d_queue}
  d_{queue} = T. 
\end{equation}

\subsection{Multi Cyclic Queuing and Forwarding}
\label{sub:mcqf}
MCQF is the further extension of CQF and 3-queue CQF, proposed to overcome the limitations of working with a single time slot in the network. Contrary to CQF, TSN 3-queue CQF, and DetNet CSQF, MCQF (shown in Fig.~\ref{fig:mcqf_overview}) works with different \texttt{Groups}, where each \texttt{Group} has two or more than two queues. Notably, each MCQF \texttt{Group} works with different time slot values denoted as ($T_{Gx}$) where $Gx$ denotes the \texttt{Group} number. The number of queues in a \texttt{Group} and the total number of \texttt{Groups} in the MCQF network is configurable. In this paper, we assume that the MCQF network has three \texttt{Groups}, denoted as $G1$, $G2$, and $G3$. There are three queues in \texttt{Group One}, which operates as a 3-queue CQF. \texttt{Group Two} and \texttt{Group Three} are each associated with two queues, thus operating as CQF. Each \texttt{Group} has a unique time slot denoted as $T_{G1}$, $T_{G2}$, $T_{G3}$, where $T_{G1}$ is the time slot value for \texttt{Group One}, $T_{G2}$ is the time slot value for \texttt{Group Two} and $T_{G3}$ is the time slot value for \texttt{Group Three}. $H$ is divided into different numbers of time slots as per different \texttt{Groups}. Fig.~\ref{fig:mcqf_hyperperiod_example} shows the number of time slots in $H$ for different \texttt{Groups}. The three \texttt{Groups} have different colors and time slot values as shown in Fig.~\ref{fig:mcqf_hyperperiod_example}. Based on the $T_{Gx}$ value, $H$ is divided into different numbers of time slots. For example as shown in Fig.~\ref{fig:mcqf_hyperperiod_example}, $H$ is divided into 16 time slots for \texttt{Group One}, 8 time slots for \texttt{Group Two}, and 4 time slots for \texttt{Group Three}. 

\noindent\textbf{WCD MCQF:} The quantitative worst case end-to-end delay of the TT flows in the MCQF network is dependent on the \texttt{Group} of the TT flow. The WCD is defined as follows:
\begin{equation}
    \mathrm{WCD_{Gx}} = f_i.\phi + (\mathrm{SW_{num}} + 1) \cdot {T_{Gx}} + (\mathrm{SW_{num}} \cdot d_{queue}) + \xi,
\end{equation}
where $Gx$ is the \texttt{Group} number, $f_i\cdot \phi$ is the offset of the flow $f_i$ in $\mu$s, $SW_{num}$ denotes the total number of switches in the routing of the TT flow, $d_{queue}$ is the queuing delay, $T_{Gx}$ is the time slot of \texttt{Group} number $x$, where $x \in \{1,2,3\}$, and $\xi$ denotes the network specific delays.

\noindent\textbf{Worst-Case Queuing Delay MCQF:} For MCQF scheduling mechanism, the worst case queuing delay ($d_{queue}$) depends on the \texttt{Group} to which the TT flow is assigned. For \texttt{Group One}, TT frames can be stored in either one of the two queues upon reception, causing the TT frames stored in the third queue to wait for one time slot before transmission. Thus, the $d_{queue}$ for \texttt{Group One} flows is one time slot. However, for MCQF \texttt{Group Two} and \texttt{Three} flows, TT frames will be transmitted in the next time slot and therefore has no waiting in the queue. The $d_{queue}$ for a TT flow ($f_i$) in MCQF network is therefore given as follows:    

\vspace{-0.1cm}
\begin{equation}
\label{eq:d_queue}
  d_{queue}(f_i) =
    \begin{cases}
      T_{G1}, \;\; \text{if} \; f_i \in G1,\\
      0, \;\;\;\;\;\; \text{otherwise }. \\
    \end{cases}       
\end{equation}

\section{System Model} 
\label{sec:system_model}
In this section, we briefly describe the system model of our paper, consisting of the network architecture, TSN switch model, and traffic model.

\subsection{Network Architecture}
\label{subsec:network_architecture} 
Our TSN network consists of switches \texttt{(SWs)} and end stations \texttt{(ESs)}. We represent the TSN network as an undirected graph denoted with \texttt{G(V, E)} where \texttt{V} denotes the nodes or vertices and \texttt{E} denotes the edges or links. \texttt{V} consists of \texttt{(ESs)} and \texttt{(SWs)} given as \texttt{V = (ESs $\cup$ SWs)} and \texttt{E} consists of all the links between the nodes in the network given as \texttt{$e_{l}$} $\in$ \texttt{E} where \texttt{l} denotes the link/edge number. The TSN network consists of periodic TT flows, all assigned a priority value of 7. The payload sizes of these flows are randomly selected between 55 and 1500 bytes (B). In our model, we assume there are eight queues~\cite{8021Q} in the egress port, with all TSN flows designated as TT flows having the highest priority. Each TT flow $(f_{i})$ is defined as a tuple of: 
\begin{equation}
    f_{i} = \langle id, src, dst, period, deadline, size, routing, \phi \rangle,
    \vspace{-0.4cm}
\end{equation}
\begin{align*}
\forall f_{i} \in F, 
\end{align*}
\noindent where $id$ is the flow number, $src$ is the source, $dst$ is the destination, $period$ is the periodicity of the flow in $\mu$s, $deadline$ is the deadline in $\mu$s, $size$ is the payload of the flow in bytes, $routing$ is the route of the TT flow from $src$ to the $dst$ and $\phi$ denotes the time offset meaning when the TT flow will be sent from the $src$ node. 

\noindent \textbf{Hypercycle:} The hypercycle $(H)$ is the scheduling cycle of the TSN network. The TT flows in the network are periodic in nature and is repeated at regular intervals. Therefore, the scheduling cycle is calculated and all the TT flows transmitted within $H$ are repeated based on its periodicity. Furthermore, $H$ remains the same for all the variants of cyclic shapers and is given as follows:
\vspace{-0.2cm}
\begin{equation}
    H = \mathrm{LCM}(f_{i}.period),\quad \forall f_i \in F.
\end{equation}

\subsection{CQF Switch Model}
In the CQF model, the TSN switches are configured with CQF scheduling mechanism with two queues: \emph{even} and \emph{odd}. The CQF mechanism operates with one time slot value denoted as $T$ and the gate operates in a ping-pong fashion in every $T$. $T$ is the time granularity where in every $T$, one queue receives and the another queue transmits. In this paper, our model gets all the input from the scheduling and configuration algorithm (for the outline refer Fig.~\ref{fig:ccnc_overview}). The offset of each TT flow is configurable in our model and given as an input from the scheduling algorithm. The offset is the time slot when the TT flows are sent from the sending node. Introducing the offset increases the total number of scheduled flows in the TSN network as shown in~\cite{itp_infocom}. For the CQF model we model the constraints as follows:

\textit{1) Offset Constraints:} The offset of each TT flow is constrained using the below equation:
\begin{equation}
    \forall f_i \in F: 0 \leq f_i\cdot \phi \leq f_i \cdot period.
\end{equation}

\textit{2) Queue resource constraint:} Since queue resources are limited in the TSN switch model, the queue resource constraint ensures that incoming flows are fully stored in their respective queues. In this paper, we use the following constraint to model our queue resources:
\vspace{-0.25cm}
\begin{equation}
    Q_{occu}(T_j) \leq Q_{len},
    \label{eq:q1}
\end{equation}
\begin{equation}
   Q_{free}(T_j) \geq f_i \cdot size,
   \label{eq:q2}
\end{equation}
\begin{equation}
    Q_{len} = Q_{occu}(T_j) + Q_{free}(T_j),
    \label{eq:q3}
\end{equation}
where $Q_{occu}(T_j)$ and $Q_{free}(T_j)$ denote the occupied and free queue sizes in a CQF-capable TSN switch at the $T_j^{th}$ time slot. $Q_{len}$ represents the total queue length of the TSN switch, which is a constant value in our framework. In our framework, $Q_{len}$ can be specified either in bytes or as the number of frames the queue can store. In the CQF network, TT frames are filtered by the \texttt{Queue Classifier} and sent to the queue currently receiving incoming frames.

\subsection{3-queue CQF Switch Model}
In the 3-queue CQF model, TSN switches are configured with a 3-queue CQF scheduling mechanism that includes three queues: \emph{even}, \emph{odd}, and \emph{tolerating}. Like CQF, the 3-queue CQF operates with a single time slot value denoted as $T$. However, in 3-queue CQF, during each time slot $T$, two queues receive frames while one queue transmits. The queue resource constraints for the 3-queue CQF model are similar to those of CQF, as described in Eq.~\ref{eq:q1}, \ref{eq:q2}, and \ref{eq:q3}. Unlike CQF, 3-queue CQF introduces a \emph{Queue ID} value, denoted as $qid$, which is generated by the scheduling algorithm for each TT flow. This $qid$ indicates which queue the TT frame will be enqueued in the 3-queue CQF switch model. In our framework, the \texttt{Queue Classifier}, as shown in Fig.\ref{fig:csqf_overview}, reads the $qid$ value and directs the TT flow to the appropriate queue in the TSN switch. The \texttt{Queue Classifier} is implemented differently for CQF and 3-queue CQF.
\subsection{MCQF Switch Model}
In our framework for the MCQF network, TT flows are initially filtered by the \texttt{Group Classifier} based on their \emph{Group} number, denoted as $gid$ in our implementation, which can be 1, 2, or 3. After filtering by the \texttt{Group Classifier}, TT flows are passed to the \texttt{Queue Classifier}. Here, the flows are sent to the correct queue within the 3-queue CQF \emph{Group} based on the $qid$ value. For the CQF \emph{Group}, the \texttt{Queue Classifier} identifies the current receiving queue and directs the TT frames to these queues. We can also use $qid$ for the MCQF \texttt{Group Two} and \texttt{Three}, however, the \texttt{Group Classifier} of CQF reads the gate status of the queues and knows which queue is currently transmitting and which one is receiving.

\textit{1) Queue resource constraint:} Since MCQF involves different time slots, the queue resource constraint is given by:
\begin{equation}
    Q_{occu}(T_{Gx}^j) \leq Q_{len},
    \label{eq:q1}
\end{equation}
\begin{equation}
    Q_{free}(T_{Gx}^j) \geq f_i \cdot size,
    \label{eq:q2}
\end{equation}
\begin{equation}
    Q_{len} = Q_{occu}(T_{Gx}^j) + Q_{free}(T_{Gx}^j),
    \label{eq:q3}
\end{equation}
where $Q_{occu}(T_{Gx}^j)$ and $Q_{free}(T_{Gx}^j)$ denote the occupied and free queue capacities in an MCQF-capable TSN switch at the $({T_{Gx}^j)}^{th}$ time slot. $T_{Gx}^j$ represents the $j$-th time slot number in $H$ of the MCQF \texttt{Group} number $Gx$. 

\begin{tcolorbox}[colback=tumColorLightBlue, colframe=black, boxrule=.2mm, boxsep=0.5mm, sharp corners=all]
    \textbf{Design Choice:} In our framework, the TSN switch queue capacity can be configured in two ways: (1) By specifying the number of frames the queue can store (an integer value), or (2) By specifying the total capacity of the queue in Bytes (B).
\end{tcolorbox}

\section{Evaluation and Discussion}  
\label{sec:results}
We implemented the CQF, 3-queue CQF, and MCQF cyclic shaper using OMNeT++ and INET4.4. To the best of our knowledge, this is the first open-source simulation tool capable of modeling all cyclic shaper variants. Additionally, it is the first time MCQF has been quantified using OMNeT++. The experimental evaluations and the benchmark of the CQF is given in our open-source code.\footnote{Github: \url{https://github.com/tum-esi/CyclicSim}}. In our framework, the queue resource can be configured either by the number of frames or by total capacity in bytes. All experiments and simulations were conducted on a Windows laptop equipped with an Intel\textregistered\ Processor Core\texttrademark\ i7-10610U CPU running at 1.80GHz, and 32 GB RAM. In this paper, for CQF and 3-queue CQF, we use $T = 50 \mu$s, and for MCQF, we use $T_{G1} = 25 \mu$s, $T_{G2} = 50 \mu$s, and $T_{G3} = 100 \mu$s unless otherwise specified in the Figures. 
\vspace{-0.1cm}
\subsection{Key Performance Indicators}
For key performance indicators (KPIs), we selected two parameters: 
\begin{enumerate}
    \item\textbf{Simulated Maximum End-to-End Delay (SMD)}: SMD denotes the time required for one TT frame to travel from the $src$ to the $dst$, expressed in microseconds ($\mu$s).
    \item\textbf{Simulated Maximum Jitter (SMJ)}: SMJ represents the maximum variation in SMD for each TT flow. 
\end{enumerate}

\subsection{Test Topologies}
\begin{figure}[!t]
    \centering
    \begin{minipage}[b]{0.19\textwidth}
        \centering
        \includegraphics[scale=0.23, trim={0.3cm 0cm 0.3cm 0.3cm}, clip]{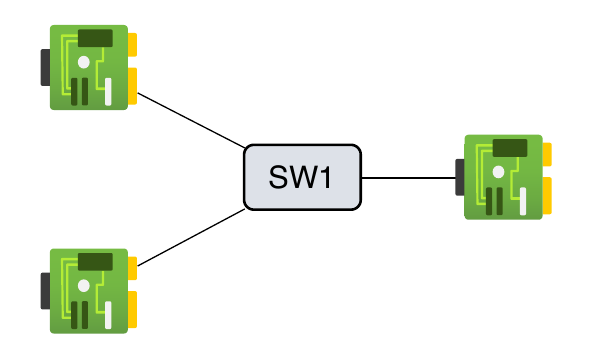}
        \subcaption[]{}
        \label{fig:one_hop_topology}
    \end{minipage}
    \hfill
    \begin{minipage}[b]{0.29\textwidth}
        \centering
        \includegraphics[scale=0.36, trim={0.8cm 0.5cm 0.7cm 0.4cm}, clip]{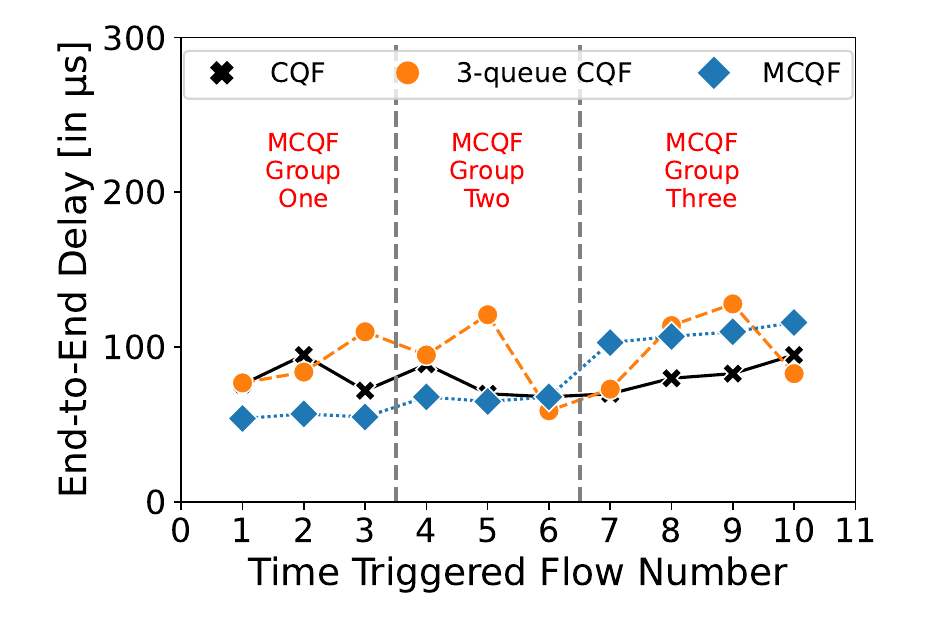}
        \subcaption[]{}
        \label{fig:one_hop_topology_results}
    \end{minipage}
    \caption{Verification using the one switch topology (Fig.~\ref{fig:one_hop_topology}) and OMNeT++ simulation results  (Fig.~\ref{fig:one_hop_topology_results}).}
    \vspace{-0.48cm}
\end{figure}

\begin{figure}[!t]{}
\begin{minipage}[b]{0.15\textwidth}
\centering
\includegraphics[scale=0.16, trim={0cm 0cm 0cm 0cm}, clip]{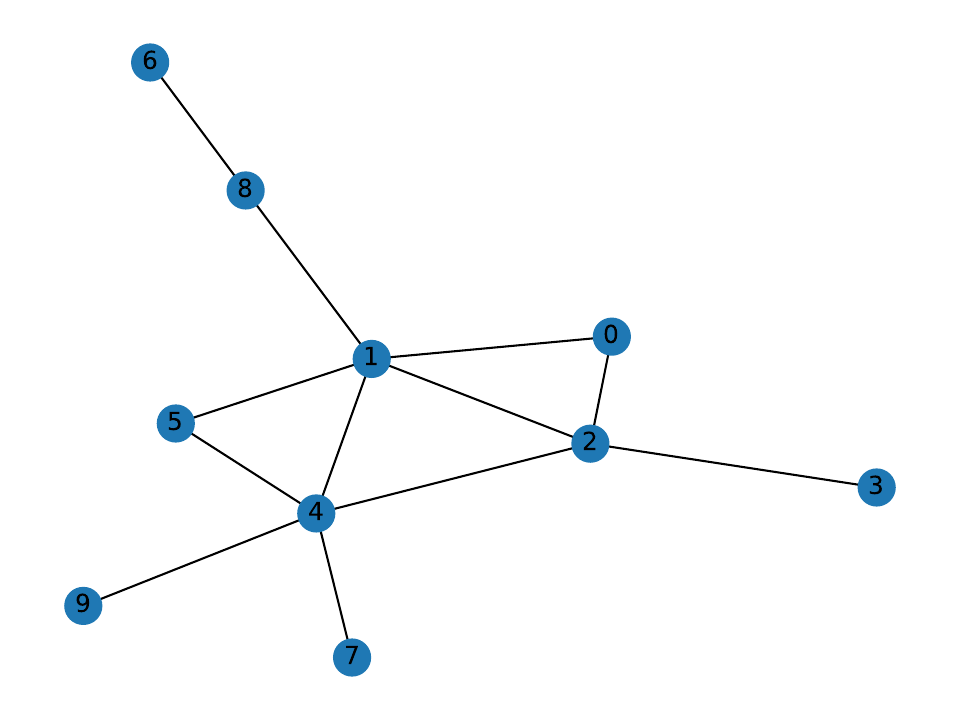}
\subcaption[]{ERG}
\label{fig:erg_topology}
\end{minipage}
\begin{minipage}[b]{0.15\textwidth}
\centering
\includegraphics[scale=0.14, trim={0cm 0cm 0cm 0cm}, clip]{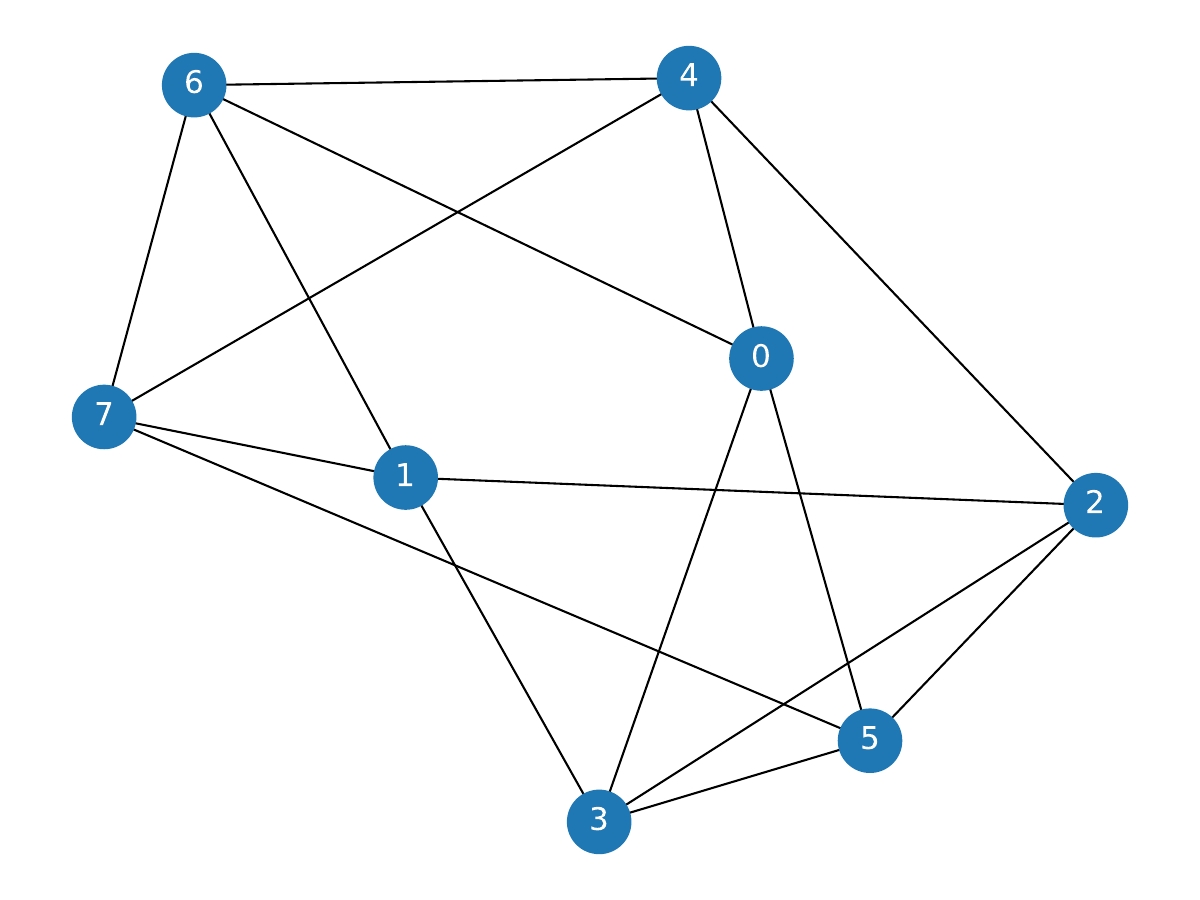}
\subcaption[]{RRG}
\label{fig:rrg_topology}
\end{minipage}
\begin{minipage}[b]{0.15\textwidth}
\centering
\includegraphics[scale=0.14, trim={0cm 0cm 0cm 0cm}, clip]{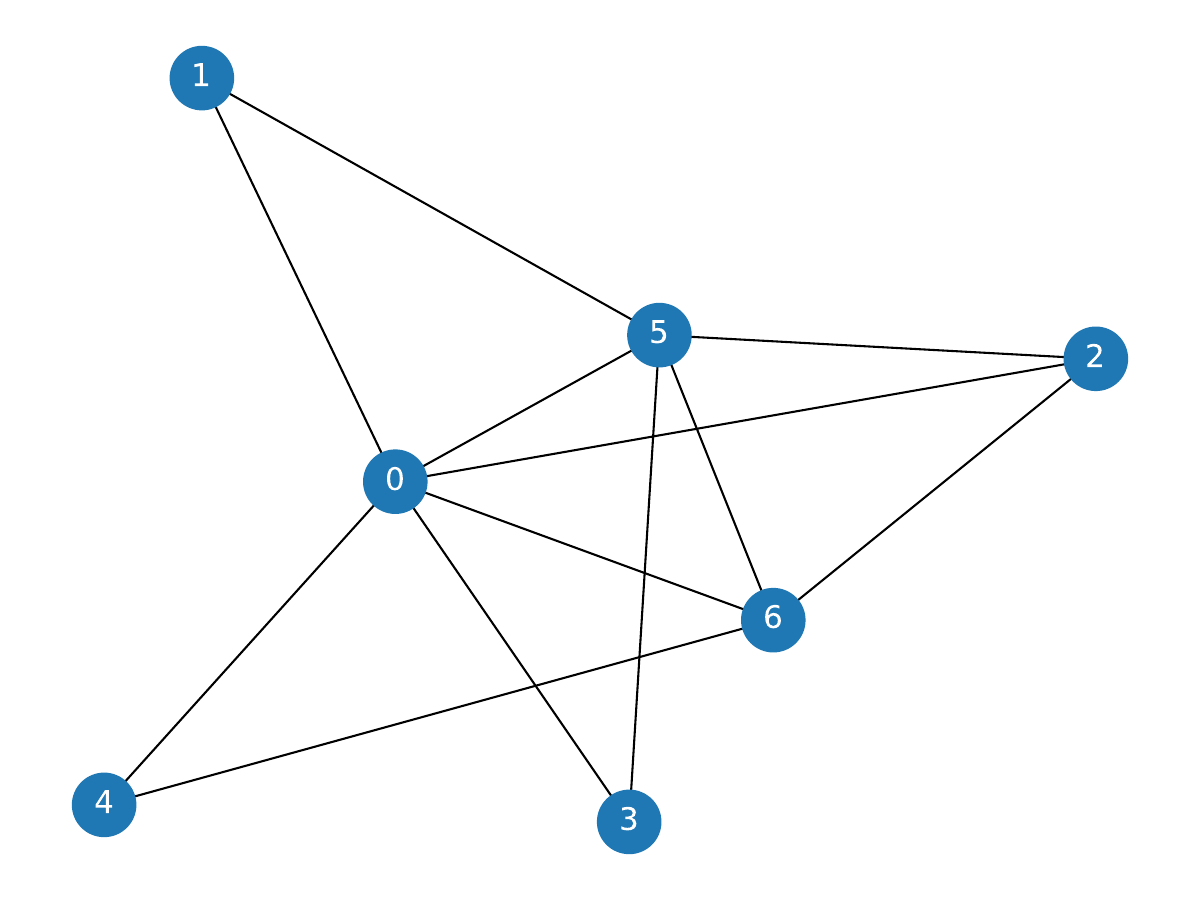}
\subcaption[]{BAG}
\label{fig:bag_topology}
\end{minipage}
\caption{Synthetic Topologies: (a) ERG topology, (b) RRG topology, and (c) BAG topology.}
\vspace{-0.38cm}
\end{figure}

\begin{figure}[!t]
\centering
\includegraphics[scale=0.2, trim={0.3cm 0cm 0.3cm 0.3cm}, clip]{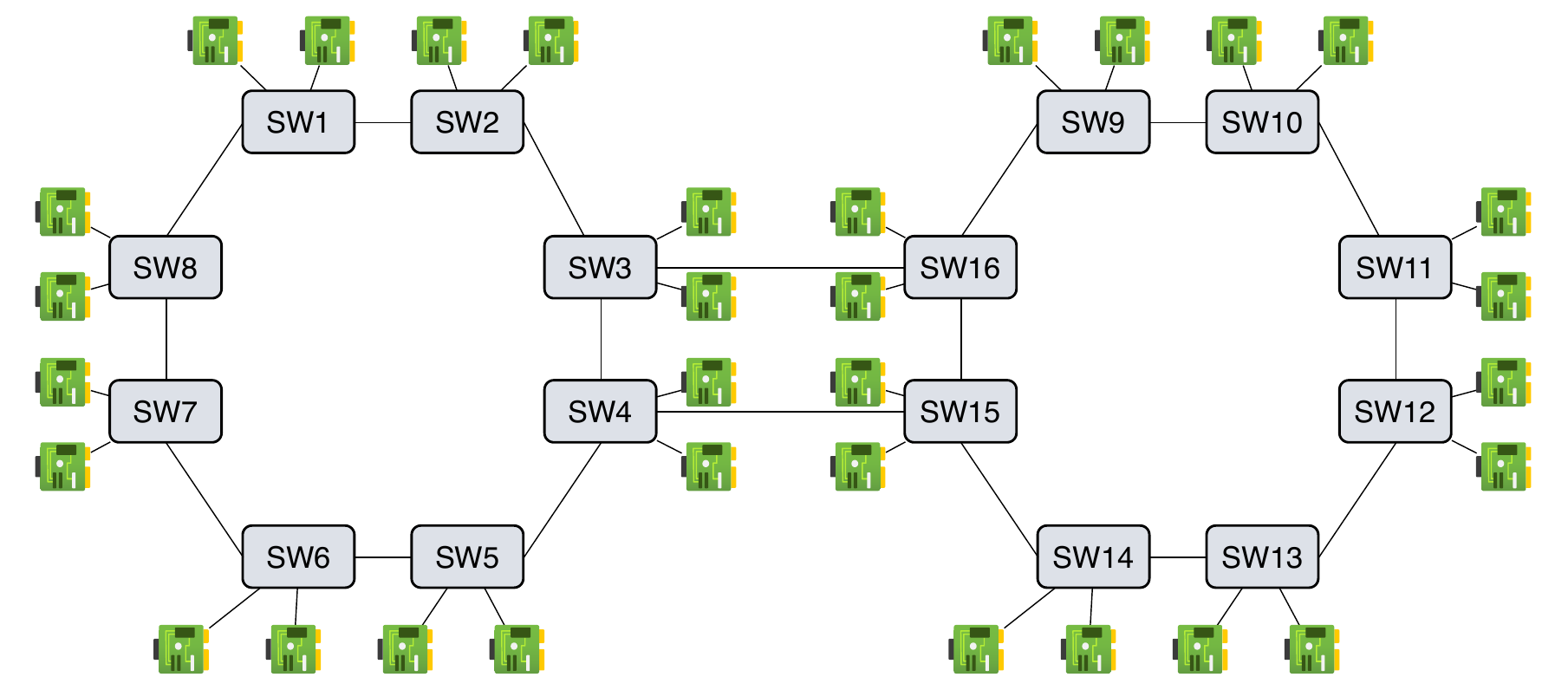}
\caption{Ring topology.}
\label{fig:ring_topology}
\vspace{-0.38cm}
\end{figure}

\begin{figure}[!t]{}
\centering
\includegraphics[scale=0.22, trim={0cm 6.9cm 2cm 0cm}, clip]{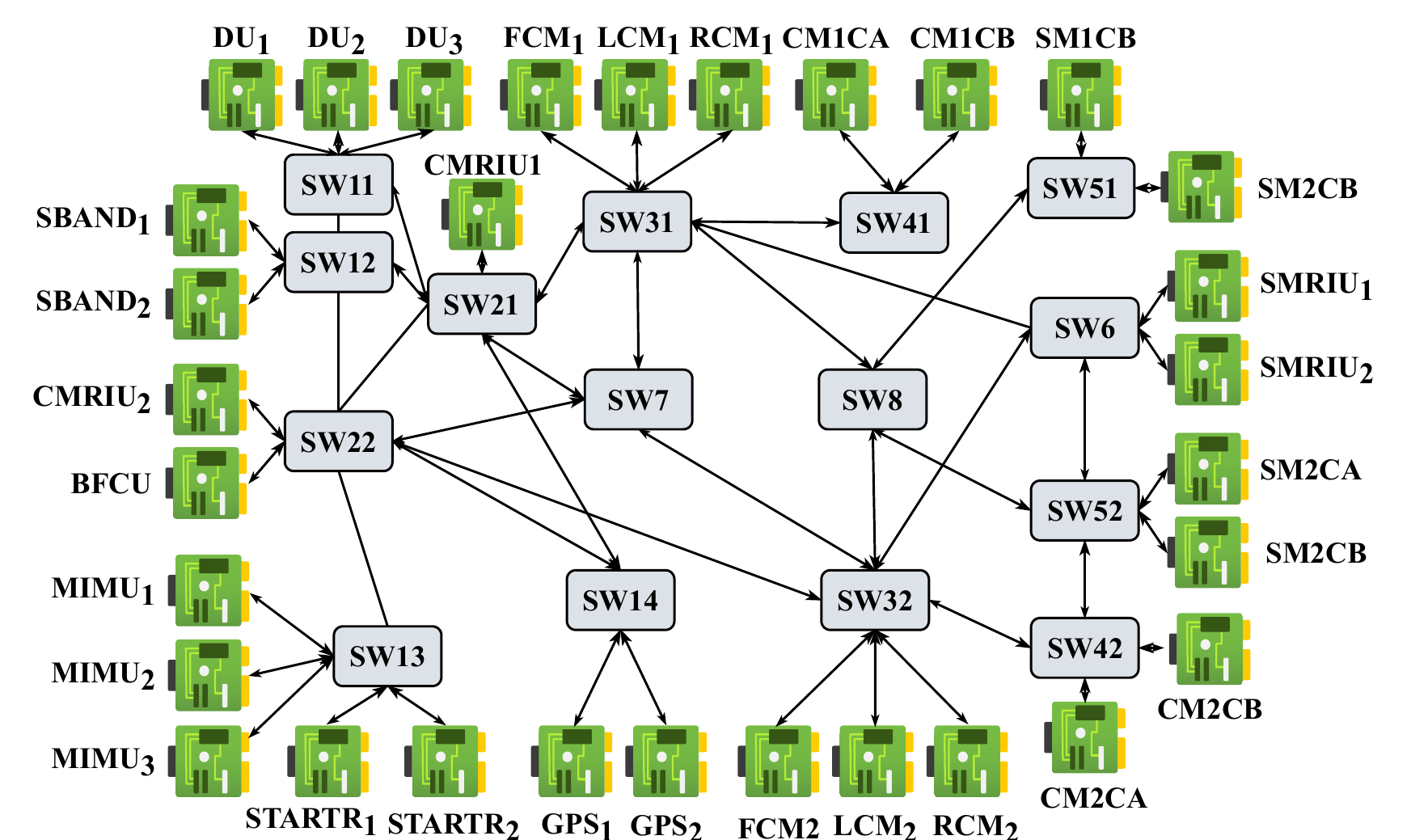}
\caption{Orion topology.}
\label{fig:orion_topology}
\vspace{-0.47cm}
\end{figure}

To illustrate the differences between various cyclic shapers, we present results for several test topologies:
\begin{enumerate}
    \item\textbf{One Switch Topology:} This simple topology, depicted in Fig.~\ref{fig:one_hop_topology}, consists of a single switch and is used for initial verification of all cyclic shapers. 
    \item\textbf{Synthetic Topologies:} For synthetic test cases, we employed different types of topologies: (1) Erdos Renyi Graph (ERG)~\cite{deepscheduler} (Fig.~\ref{fig:erg_topology}), (2) Random Regular Graph (RRG)~\cite{deepscheduler} (Fig.~\ref{fig:rrg_topology}), and (3) Barabasi-Albert Graph (BAG)~\cite{deepscheduler} (Fig.~\ref{fig:bag_topology}). 
    \item\textbf{Industrial Topology:} Ring topologies are commonly used in industrial applications. Therefore, we used a Ring topology, as shown in Fig.~\ref{fig:ring_topology}. 
    \item\textbf{Realistic Topology:} For a realistic test case, we utilized the Orion Crew Exploration Vehicle (CEV) topology~\cite{rubi_noms}, depicted in Fig.~\ref{fig:orion_topology}. 
\end{enumerate}

\subsection{One Switch Topology} 
\label{sec:one_hop}
\begin{figure*}[!t]{}
\begin{minipage}[b]{0.33\textwidth}
\centering
\includegraphics[scale=0.35, trim={0.7cm 0.5cm 1cm 0.4cm}, clip]{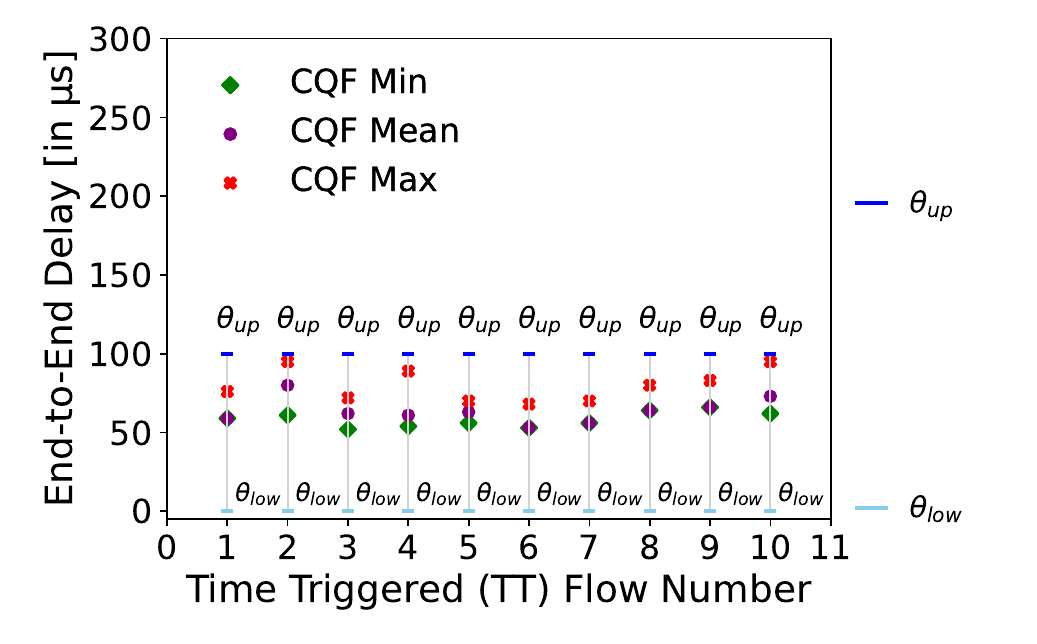}
\subcaption[]{CQF.}
\label{fig:upper_lower_cqf_one_hop}
\end{minipage}
\begin{minipage}[b]{0.33\textwidth}
\centering
\includegraphics[scale=0.38, trim={0.4cm 0.5cm 0.8cm 0.4cm}, clip]{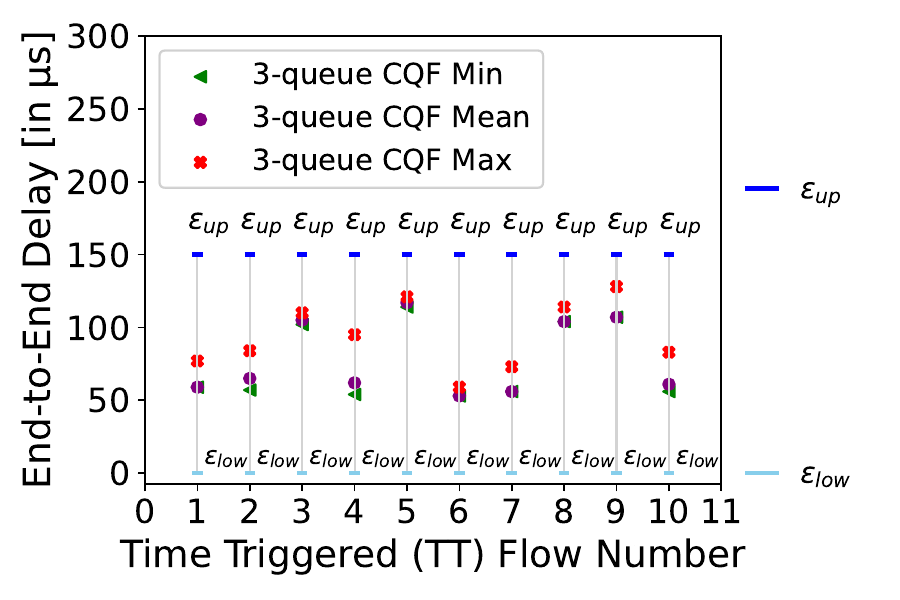}
\subcaption[]{3-Queue CQF.}
\label{fig:upper_lower_csqf_one_hop}
\end{minipage}
\begin{minipage}[b]{0.33\textwidth}
\centering
\includegraphics[scale=0.36, trim={0.7cm 0.5cm 1cm 0.4cm}, clip]{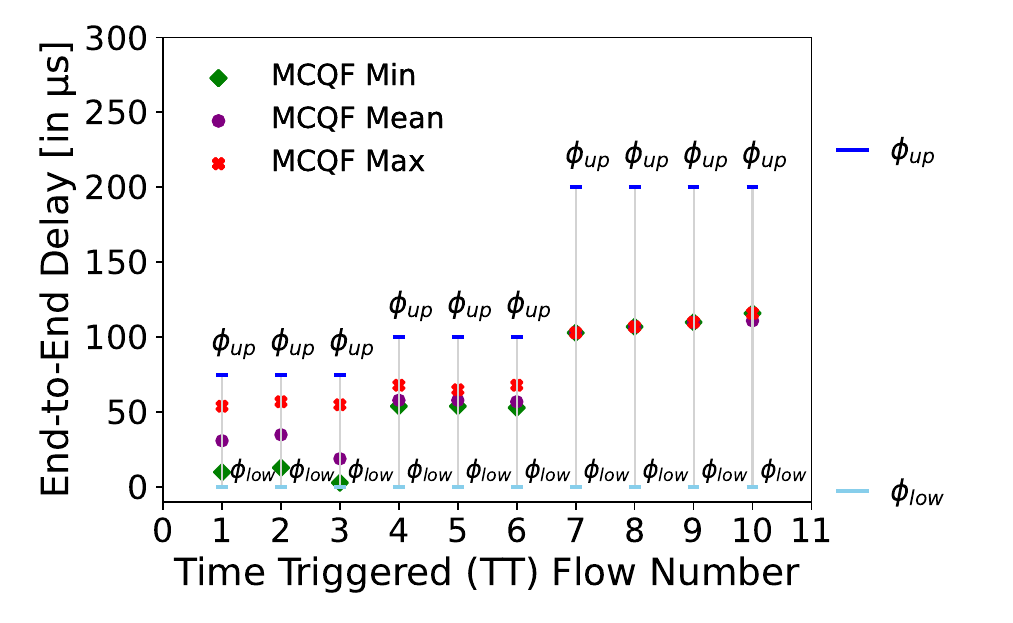}
\subcaption[]{MCQF.}
\label{fig:upper_lower_mcqf_one_hop}
\end{minipage}
\caption{End-to-End Delay: quantitative upper bound, lower bound and simulative delay for one switch topology (Fig.~\ref{fig:one_hop_topology}).}
\vspace{-0.28cm}
\end{figure*}

\begin{figure*}[!t]{}
\begin{minipage}[b]{0.33\textwidth}
\centering
\includegraphics[scale=0.34, trim={0.6cm 0.5cm 0.5cm 0.4cm}, clip]{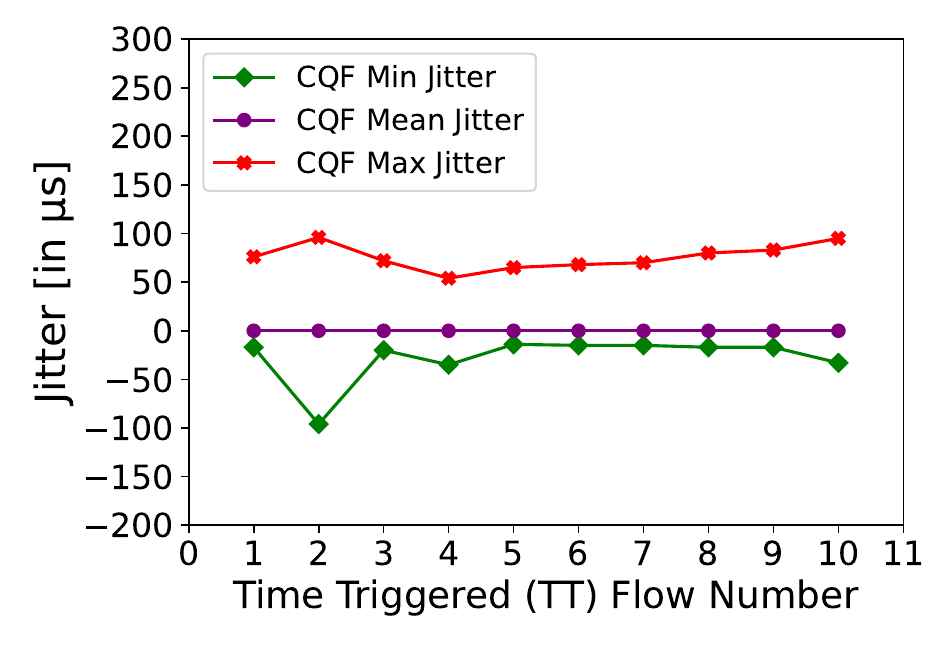}
\subcaption[]{CQF.}
\label{fig:upper_lower_cqf_one_hop_jitter}
\end{minipage}
\begin{minipage}[b]{0.33\textwidth}
\centering
\includegraphics[scale=0.37, trim={0.5cm 0.5cm 0.5cm 0.4cm}, clip]{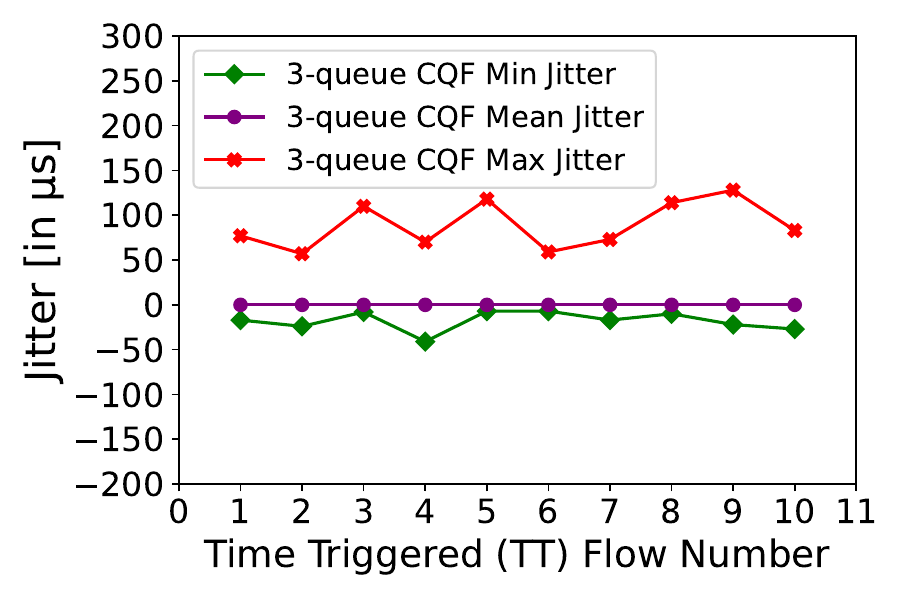}
\subcaption[]{3-Queue CQF.}
\label{fig:upper_lower_csqf_one_hop_jitter}
\end{minipage}
\begin{minipage}[b]{0.33\textwidth}
\centering
\includegraphics[scale=0.37, trim={0.5cm 0.5cm 0.5cm 0.4cm}, clip]{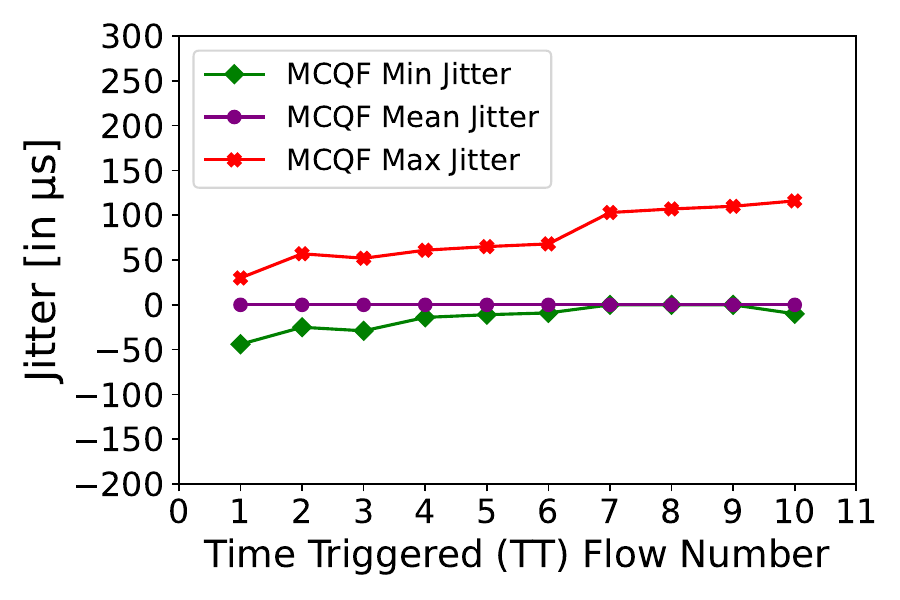}
\subcaption[]{MCQF.}
\label{fig:upper_lower_mcqf_one_hop_jitter}
\end{minipage}
\caption{Jitter comparison for one switch topology (Fig.~\ref{fig:one_hop_topology}).}
\vspace{-0.38cm}
\end{figure*}

In this sub-section, we compare the performance of different cyclic shapers using the one switch topology shown in Fig.~\ref{fig:one_hop_topology}. We simulate this topology to validate our framework and ensure accurate results for the cyclic shapers. The simulation results, presented in Fig.~\ref{fig:one_hop_topology_results}, show the maximum end-to-end delay for the three cyclic shapers categorized by the MCQF \texttt{Group} number. The results indicate that the maximum delay for MCQF \texttt{Group One} is lower than that of CQF and 3-queue CQF, primarily because the time slot ($T_{G1}$) for MCQF $G1$ is smaller than the $T$ value used for CQF and 3-queue CQF. For validation, we include the theoretical upper and lower bounds: $\theta_{up}$ and $\theta_{low}$ denotes the theoretical upper and lower bounds for CQF, $\epsilon_{up}$ and $\epsilon_{low}$ for 3-queue CQF, and $\phi_{up}$ and $\phi_{low}$ for MCQF in the same graph as shown in Fig.~\ref{fig:upper_lower_cqf_one_hop}, \ref{fig:upper_lower_csqf_one_hop}, and \ref{fig:upper_lower_mcqf_one_hop}. This allows us to demonstrate that the minimum, mean, and maximum simulated delays remain within these bounds. The results confirm that the simulated delays are consistent with the theoretical expectations, thereby validating the accuracy of our simulation framework. Fig.~\ref{fig:upper_lower_cqf_one_hop_jitter}, \ref{fig:upper_lower_csqf_one_hop_jitter}, and \ref{fig:upper_lower_mcqf_one_hop_jitter} illustrate the jitter for CQF, 3-queue CQF, and MCQF, respectively. Notably, the jitter increases for MCQF as the flow transitions to \texttt{Group Three}, indicating a higher variation in delay for these flows.

\vspace{-0.1cm}
\begin{tcolorbox}[colback=tumColorLightBlue, colframe=black, boxrule=.3mm, boxsep=0.3mm, sharp corners=all]
    \textbf{Finding 1:} CQF provides consistent QoS across all traffic types and performs well in scenarios with a single set of timing requirements.
\end{tcolorbox}

\begin{figure*}[!t]{}
\begin{minipage}[b]{0.44\textwidth}
\centering
\includegraphics[scale=0.35, trim={0.5cm 0.5cm 0.4cm 0cm}, clip]{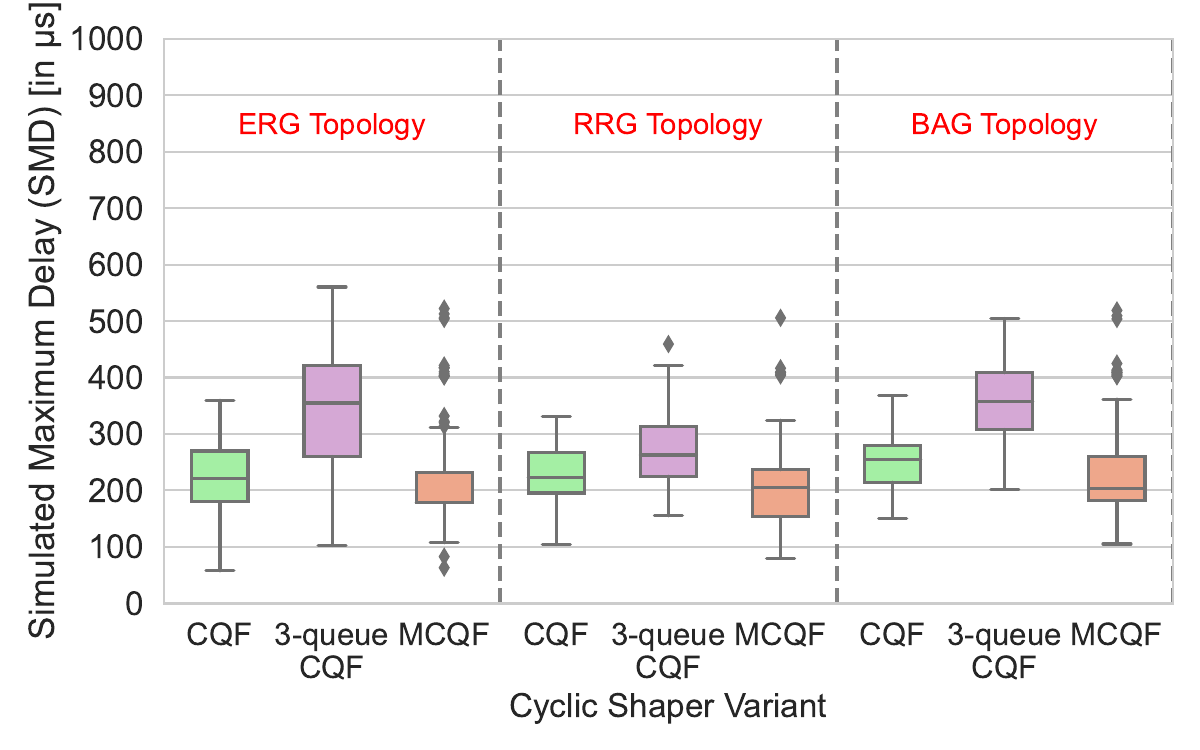}
\subcaption[]{Simulated Maximum End-to-End Delay}
\label{fig:all_delay}
\end{minipage}
\hfill
\begin{minipage}[b]{0.44\textwidth}
\centering
\includegraphics[scale=0.35, trim={0.5cm 0.5cm 0.3cm 0cm}, clip]{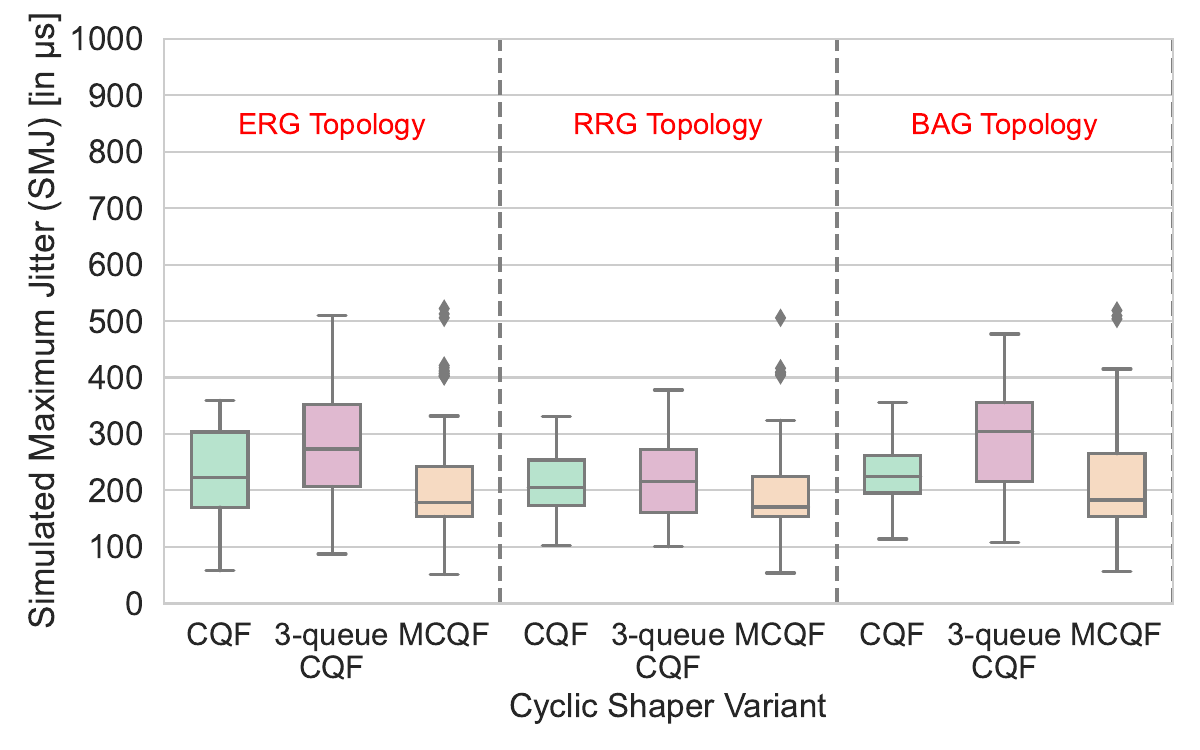}
\subcaption[]{Simulated Maximum Jitter}
\label{fig:all_jitter}
\end{minipage}
\caption{End-to-End Delay and Jitter comparison for CQF, 3-queue CQF and MCQF for different topologies.}
\vspace{-0.48cm}
\end{figure*}

\vspace{-0.3cm}
\subsection{Synthetic Topologies}
\label{sec:synthetic}
Fig.~\ref{fig:all_delay} and \ref{fig:all_jitter} present the delay and jitter for three different topologies: ERG, RRG, and BAG. The outliers showing larger delays for MCQF correspond to the flows in \texttt{Group Three}. Notably, MCQF consistently exhibits lower delays compared to CQF and 3-queue CQF, primarily because the time slot value for \texttt{Group One} in MCQF is smaller. Interestingly, 3-queue CQF does not demonstrate any delay advantage over CQF. In fact, the delay for 3-queue CQF is larger, attributed to the \emph{tolerating queue}. While 3-queue CQF can schedule a higher number of flows due to the additional queue, this comes at the cost of increased delay. Additionally, the jitter is higher for both 3-queue CQF and MCQF, with the TT flows in MCQF's \texttt{Group Three} experiencing the highest jitter, as the time slot configuration directly influences jitter values. However, this elevated jitter is specific to \texttt{Group Three} flows. In contrast, \texttt{Group One} and \texttt{Group Two} flows in MCQF experience smaller delays and lower jitter values, demonstrating the time slot's significant impact on performance across different flow \texttt{Groups}.

\subsection{Industrial Ring Topology}
\label{sec:industrial}
\begin{figure}[!t]{}
\centering
\includegraphics[scale=0.286, trim={0.3cm 0.3cm 0.3cm 0.1cm}, clip]{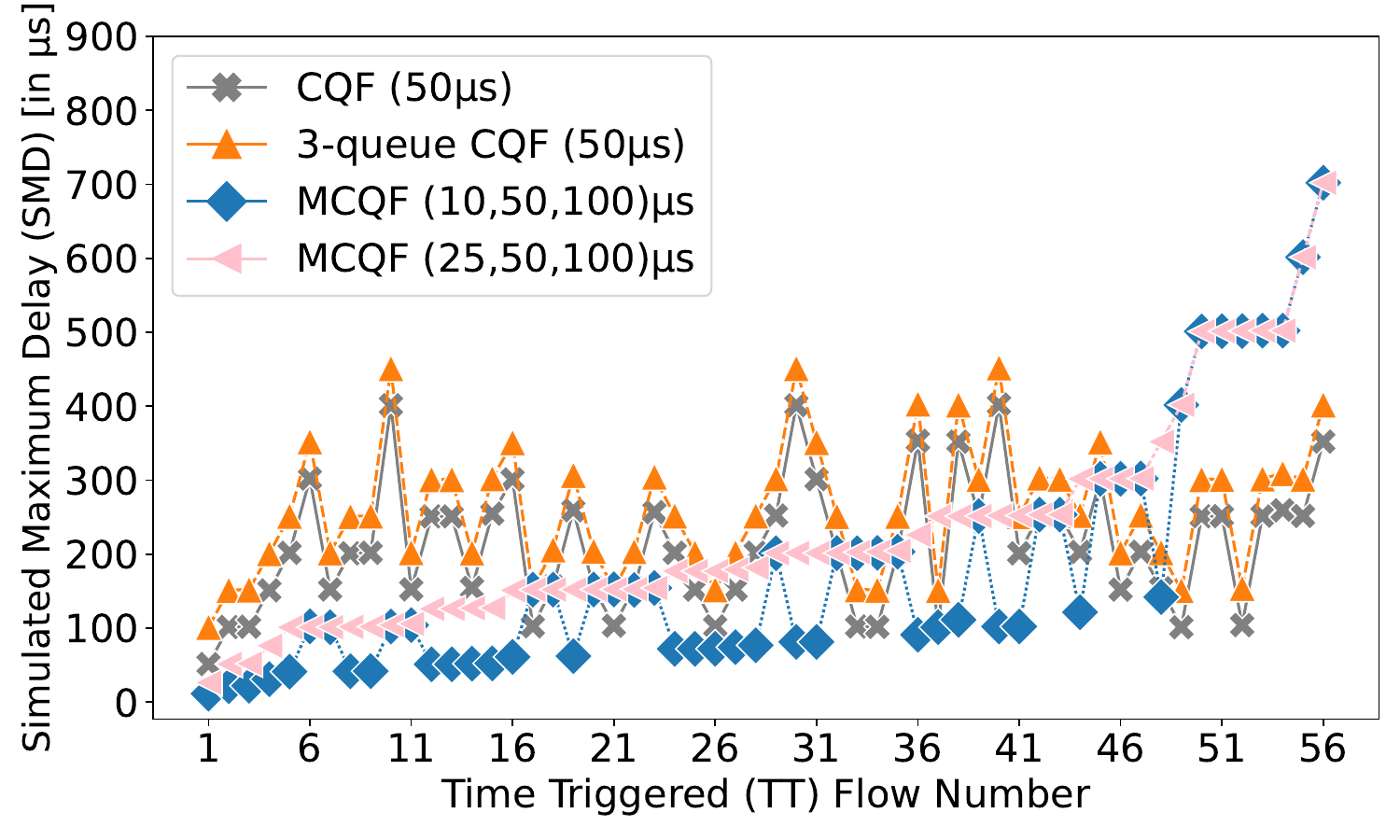}
\caption{CQF, 3-queue CQF, MCQF: End-to-End Delay comparison for Ring topology (Fig.~\ref{fig:ring_topology}).}
\label{fig:all_delay_ring}
\vspace{-0.5cm}
\end{figure}

Fig.~\ref{fig:all_delay_ring} illustrates the end-to-end delay of the cyclic shapers for the ring topology (refer Fig.~\ref{fig:ring_topology}). The flows on the x-axis are sorted from the smallest to the largest $gid$ value. For MCQF, the delay is smallest for the \texttt{Group One} flows, but it increases significantly for the \texttt{Group Three} flows due to the larger time slot value. In contrast, CQF provides consistent delay and QoS for all flows, with the primary factor being the number of switches involved in routing the TT flow. Meanwhile, 3-queue CQF exhibits larger delays compared to both CQF and MCQF for \texttt{Group One} and \texttt{Group Two} flows. MCQF is clearly well-suited for accommodating both hard real-time and soft real-time traffic types. Furthermore, based on our experiments, the 3-queue CQF for TSN does not demonstrate any significant advantage over CQF and MCQF.

\begin{tcolorbox}[colback=tumColorLightBlue, colframe=black, boxrule=.3mm, boxsep=0.3mm, sharp corners=all]
    \textbf{Finding 2:} 3-queue CQF leads a larger delay and jitter than CQF due to the \emph{tolerating queue}. 3-queue CQF is not highly advantageous in TSN as the propagation delays and time synchronization error is not significant in TSN as compared to WAN.
\end{tcolorbox}
\vspace{-0.16cm}

\begin{tcolorbox}[colback=tumColorLightBlue, colframe=black, boxrule=.3mm, boxsep=0.3mm, sharp corners=all]
    \textbf{Finding 3:} MCQF demonstrates its superiority by effectively supporting diverse timing requirements accommodating very small to very large timing requirements.
\end{tcolorbox}
\begin{tcolorbox}[colback=tumColorLightBlue, colframe=black, boxrule=.3mm, boxsep=0.3mm, sharp corners=all]
    \textbf{Finding 4:} Most of the configuration and scheduling algorithm do not consider the propagation delay, switching delay, time synchronization error etc. The simulation results shows that these delays cannot be ignored and should be included into the optimization phase.
\end{tcolorbox}
\renewcommand{\baselinestretch}{0.95}
\subsection{Realistic Topology}
\label{sec:realistic}
\begin{figure}[h]{}
\centering
\includegraphics[scale=0.29, trim={0.3cm 0.3cm 0.3cm 0.3cm}, clip]{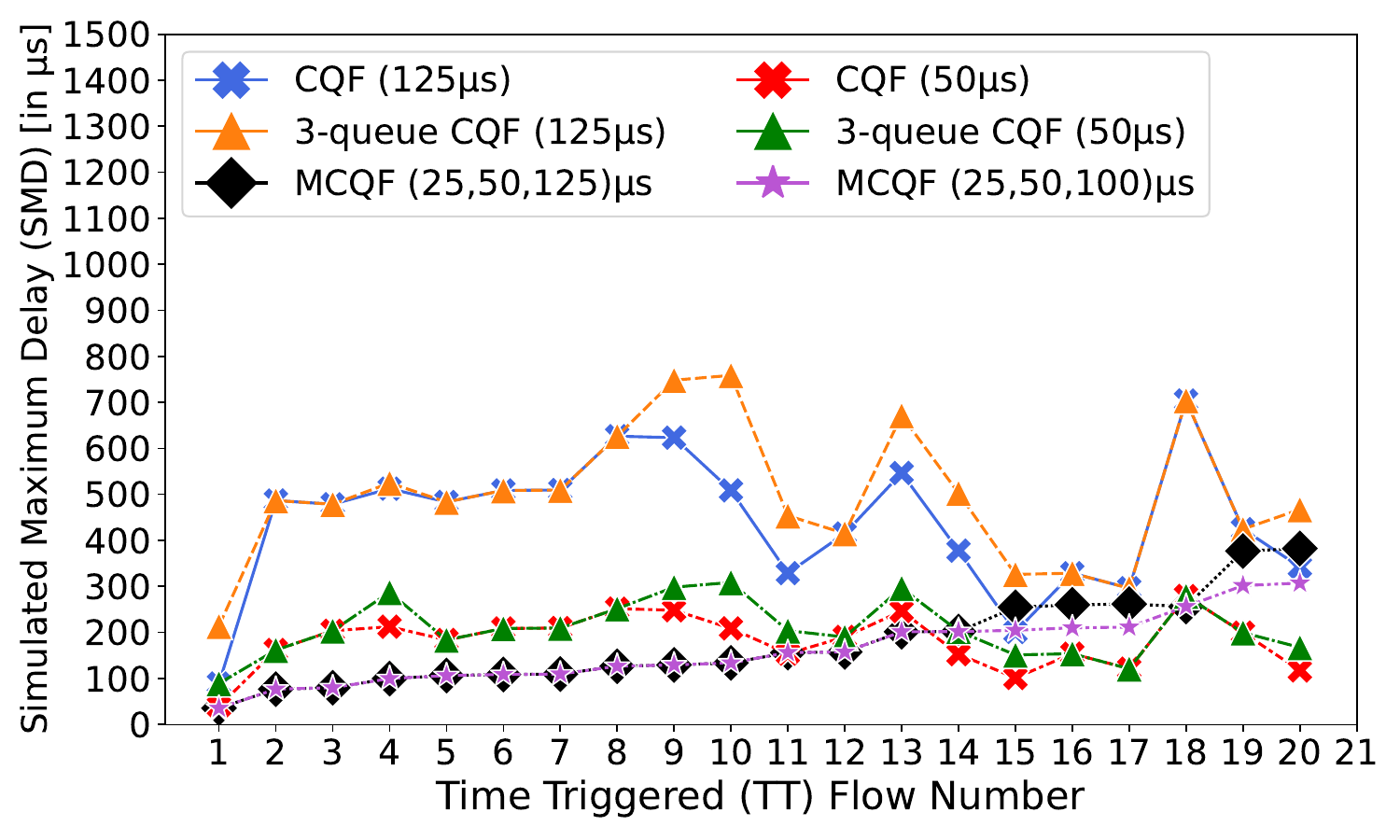}
\caption{CQF, 3-queue CQF, MCQF: End-to-End Delay comparison for Orion topology (Fig.~\ref{fig:orion_topology}).}
\label{fig:all_delay_orion}
\vspace{-0.18cm}
\end{figure}

Fig.~\ref{fig:all_delay_orion} shows the delay performance of the Orion topology (refer Fig.~\ref{fig:orion_topology}). MCQF outperforms CQF and 3-queue CQF for all the three \texttt{Groups}. As the network load is very low in the Orion topology, the gate \texttt{opening} and \texttt{closing} is not very frequent and there are no competing flows. Therefore, MCQF \texttt{Group Three} flows exhibit good performance and MCQF totally outperforms CQF and 3-queue CQF. 

\vspace{-0.1cm}
\begin{tcolorbox}[colback=tumColorLightBlue, colframe=black, boxrule=.3mm, boxsep=0.3mm, sharp corners=all]
    \textbf{Finding 5:} CSQF in Layer 3 differs from the TSN 3-queue CQF. Therefore, it is crucial to use accurate terminology to avoid confusion.
\end{tcolorbox}

\begin{tcolorbox}[colback=tumColorLightBlue, colframe=black, boxrule=.2mm, boxsep=0.2mm, sharp corners=all]
    \textbf{Finding 6:} TSN switch queue capacity directly affects the overall performance of the TT flows in the network.
\end{tcolorbox}

\begin{tcolorbox}[colback=tumColorLightBlue, colframe=black, boxrule=.2mm, boxsep=0.2mm, sharp corners=all]
    \textbf{Finding 7:} MCQF network requires the $\textbf{gid}$ and $\textbf{qid}$ information. Adding this information directly to the Ethernet header would increase its header size. A more efficient approach, as used in our implementation, is to utilize tags for the TT flow, which simplifies the process and optimizes the implementation.
\end{tcolorbox}

\begin{table}[t!]
    \begin{threeparttable}
    	\caption{Cyclic Shapers Simulation Comparison}
        \label{tab:related_work_table}
        \renewcommand{\arraystretch}{1.2} 
        \setlength{\tabcolsep}{4pt} 
        \begin{tabular}{l c c c c}
    		\toprule
    		\multirow{2}{*}{\rotatebox[origin=c]{0}{\textbf{Related Work}}} &
    		\multirow{2}{*}{\rotatebox[origin=c]{0}{\textbf{CQF}}} &
            \multirow{2}{*}{\rotatebox[origin=c]{0}{\textbf{CSQF (3-queue CQF)}}} &
            \multirow{2}{*}{\rotatebox[origin=c]{0}{\textbf{MCQF}}} &
            \multirow{2}{*}{\rotatebox[origin=c]{0}{\textbf{Open Source}}} \\
    		&
    		& 
            & \\
    		\midrule
        	\cite{qch_wang_performance} & \harveyBallFull & \harveyBallNone & \harveyBallNone & \harveyBallNone \\
        	\cite{assessment_qch_leonardi} & \harveyBallFull &  \harveyBallNone & \harveyBallNone & \harveyBallNone \\
            \cite{sensors_cqf} & \harveyBallFull &  \harveyBallNone & \harveyBallNone & \harveyBallNone \\
        	\cite{iecon_cqf} & \harveyBallFull &  \harveyBallNone & \harveyBallNone & \harveyBallNone \\
        	INET4.4 & \harveyBallNone & \harveyBallNone & \harveyBallNone & \harveyBallFull \\
    		\textbf{CyclicSim} & \harveyBallFull & \harveyBallFull & \harveyBallFull & \harveyBallFull \\
    		\bottomrule
    		\multicolumn{5}{l}{  \harveyBallNone\ no or n/a, \harveyBallHalf\ partially, \harveyBallFull\ yes}
    	\end{tabular}
    \end{threeparttable}
    \vspace{-0.58cm}
\end{table}

\section{Related Work}
\label{sec:related_work}
In recent years, extensive research~\cite{ahmed_access, rubi_rtcsa, anna_fp, rubi_noms, qch_wang_performance, assessment_qch_leonardi,reusch2023configuration} has analyzed the performance of various TSN shaping mechanisms. Among these, OMNeT++ is the most widely used tool, as highlighted in \cite{arxiv_sch_survey}. Simulation methods provide valuable insights into real-world scenarios and help analyze network performance. Moreover, simulation is often the most practical approach for verifying network performance when hardware resources are either limited or unavailable. Nasrallah et al. evaluated the average and the maximum delay of TAS in industrial ring topologies for sporadic and periodic traffic types and also evaluated the Asynchronous Traffic Shaper (ATS) using OMNeT++. Arestova et al. compared TAS and frame preemption (FP) with strict priority (SP) in \cite{anna_fp}. Debnath et al. further showed the individual and the combined shaper simulative analysis of TSN shapers in \cite{rubi_rtcsa} and \cite{rubi_noms}. However, none of these studies cover different variants of CQF.

Wang et al. in \cite{qch_wang_performance} provided a performance evaluation of IEEE 802.1Qch CQF and compared it with IEEE 802.1Qbv TAS~\cite{8021Qbv}. Their goal was to compare the performance of CQF against TAS. The paper highlighted that while the end-to-end delay of CQF is bounded, CQF provides a larger delay than TAS and fails to offer the same level of determinism due to jitter. Leonardi et al. in \cite{assessment_qch_leonardi} further evaluated the performance of CQF and compared it against CBS in an automotive scenario. They used OMNeT++ to simulate the CQF and CBS networks. Their paper shows the performance of the TSN traffic when CQF is used instead of CBS. Luo et al. in \cite{iecon_cqf} used the CoRE4INET framework and performed the evaluation of CQF and compared it against TAS. Although, CoRE4INET framework is developed based on the older version of INET and therefore lacks many complex features which is required by TSN. Furthermore, both of the works have not evaluated the CSQF and the MCQF network in their work. To the best of our knowledge, currently there is no available simulative performance evaluation of CQF and its variants. Table~\ref{tab:related_work_table} presents the comparison between the existing cyclic shaper simulation works with our framework: CyclicSim.

\vspace{-0.3cm}
\section{Conclusion}
\label{sec:conclusion}
In this paper, we present an open-source framework for simulating the different variants of cyclic shapers in TSN. We provide a comprehensive performance evaluation of CQF, CSQF (3-queue CQF), and MCQF across one-hop, synthetic, industrial, and realistic test cases. Our evaluation reveals that the end-to-end delay is higher for the CSQF (3-queue CQF) compared to CQF, primarily due to the \emph{tolerating queue}. Both CQF and CSQF (3-queue CQF) operate with a single time slot, limiting their ability to support diverse traffic types with varying timing requirements. In TSN networks, the \emph{tolerating queue} is often underutilized since propagation delay and time synchronization errors are typically negligible or constant. In contrast, MCQF supports a wide range of timing requirements, offering a diverse range of QoS options. Therefore, selecting the appropriate cyclic shaper is crucial based on the specific application. In future work, we plan to evaluate the combined shaping mechanisms of cyclic shapers with TAS, CBS, ATS, SP and FP.

\bibliographystyle{IEEEtran}

\renewcommand{\baselinestretch}{0.87}
\bibliography{reference}

\end{document}